\documentclass[twocolumn]{aastex631}

\newcommand{\arcs}{$^{\rm \prime\prime}$}
\newcommand{\msun}{M$_{\odot}$}

\usepackage{soul}
\shorttitle{I~Zw~81}
\shortauthors{Pandey et al.}
\graphicspath{{./}{figures/}}
\begin{document}

\title{Central star formation in an early-type galaxy I~Zw~81 in the Bootes void}
\footnote{Released on March, 1st, 2021}

\correspondingauthor{Divya Pandey}
\email{divya\_pandey@nitrkl.ac.in}
%in square bracket write orchid id
\author{Divya Pandey}
\affiliation{Department of Physics and Astronomy,
National Institute of Technology,
Rourkela, Odisha 769 008, India}

\author{Kanak Saha}
\affiliation{Inter-University Centre for Astronomy \& Astrophysics,
Postbag 4, Ganeshkhind, Pune 411 007, India
}

%\nocollaboration{1}

\author{Ananta C. Pradhan}
\affiliation{Department of Physics and Astronomy,
National Institute of Technology,
Rourkela, Odisha 769 008, India}

\author{Sugata Kaviraj}
\affiliation{Centre for Astrophysics Research, Department of Physics, Astronomy and Mathematics, University of Hertfordshire, Hatfield AL10 9AB, UK}


\begin{abstract}
The origin of star-formation in customarily passively evolving early-type massive galaxies is poorly understood. We present a case study of a massive galaxy, I~Zw~81, inside the Bootes void. The void galaxy is known to host an active galactic nuclei (AGN). Our detailed 2D decomposition of the surface brightness distribution in the $Canada$ $France$ $Hawaii$ $Telescope$ ($CFHT$) g- and r-bands revealed multiple structural components such as a {nuclear point source}, a bar, a ring, and an inner exponential disk followed by an outer low surface brightness (LSB) disk. I~Zw~81 turns out to be a disk-dominated galaxy with lenticular morphology. The modelling of the multi-wavelength spectral energy distribution (SED) shows that the galaxy is star-forming (SF), and belongs to the blue cloud. We find that the optical (g$-$r) color of the bar is bluer than the disks, and the far- and near-ultraviolet emission inside the galaxy observed with Ultraviolet Imaging Telescope (UVIT) onboard {\em AstroSat} is concentrated in the central few kpc region enclosing the bar. The strong bar might be playing a pivotal role in driving the gas inflow and causing SF activity in tandem with the minor merger-like interactions as evident from the deep $CFHT$ data. The low-luminosity AGN is insufficient to quench the central SF. The results are peculiar from the standpoint of a massive barred lenticular galaxy. 

%Further observation with Integral Field Units (IFU) would be essential to understand how such massive void galaxies are forming stars and evolving.

%Structural evidences confirm that the galaxy has evolved secularly with no signs of merger-like interaction.  The low-luminosity AGN is insufficient to quench the central SF. The results are peculiar to find from the standpoint of a barred massive lenticular galaxy. Further observation with IFU would be essential to understand how such massive void galaxies form and evolve. 

\end{abstract}



\section{Introduction}
\label{sec:intro}

Galaxy evolution is mainly influenced by their environment \citep{1980Dressler, 2007Cooper} and stellar mass \citep{2003KauffmannM,2015Alpaslan}. The physical mechanisms responsible for the growth of a galaxy due to the environment ({\it galaxy nurture}) include galaxy interactions, gas accretion and feedback \citep[][]{Kauffmann2004,gasaccretion2016}, ram pressure stripping \citep{1972Gunn}, galaxy harassment \citep{1976Richstone,1996Galhar} and halo quenching \citep{2006Dekelandb}. On the other hand, the stellar mass affects galaxy evolution through various mechanisms such as  morphological quenching \citep{2009Martig}, bar quenching \citep[][]{2018Khoperskov}, gravitational quenching \citep{2014Genzel}, and active galactic nuclei (AGN) feedback \citep{2006Croton}. Many of these processes may act on a galaxy simultaneously throughout its lifetime, which makes it difficult to identify the effect of an individual mechanism on a galaxy, especially in case of an over-dense environment. 

\par

Low-density environment provides a unique setting to study galaxy evolution where the environmental effects can be largely excluded. Voids are the large under-dense regions in the cosmic web having an astoundingly low mean galaxy-density \citep{pan2012}. The low-density environment of the void affects some of the properties of the void galaxies, e.g., these galaxies possess higher specific star formation rates (sSFRs) than their counterparts in the dense environment \citep{1999Grogin}. Generally, void galaxies are star-forming low mass-systems with late-type morphology \citep{ 2004Rojas,2002Cruzen, 2012Kreckel, 2021Pandey}. A fraction of them are found to be passive, early-type, interacting, or with active nuclei at the center \citep{2008Constantin,2017Beygu}. While a majority of the studies probing void galaxies focus on low-to-moderately massive void galaxies ($\lesssim 10^{10}$ \msun) \citep[see][]{1999Grogin,2004Rojas,2011PT,2012Kreckel}, spectro-photometric properties of void galaxies falling in the higher stellar mass regime are comparatively less discussed. One of the most intuitive reasons for the under representation of massive void galaxies in the literature is their sheer scarcity in the cosmic web \citep{1999Grogin,2011Tempel}.

It is well established that the environment plays a decisive role in quenching the star-formation of low mass satellite galaxies \citep{Bluck2020}. At the same time, the situation becomes unclear in case of high mass central galaxies. \citet[][]{2003KauffmannM} have shown that the fraction of quenched galaxies increases after a threshold mass $\simeq$ 3 $\times$ 10$^{10}$ \msun. Agreeably various mass quenching mechanisms viz. AGN quenching, bar quenching, virial shock heating, etc., become active at high stellar masses \citep{2021Zhang}. However, these mechanisms seem to have both positive and negative evidences, e.g., \citet{2012Maiolino} reported that AGN driven outflows at high redshifts can clear the gas out of a galaxy leading to a rapid shut down of star formation. In contrast, the analyses done by  \citet{2018Shangguan,2020Shangguan} do not support AGN feedback quenching. Therefore, the entire process of galaxy quenching
is complex and demands further study. 

\citet{2021Pandey} studied the properties of star-forming void galaxies based on deep ultraviolet (UV) survey in the Bootes void using Ultra-Violet Imaging Telescope (UVIT). In this work, I~Zw~81 was identified as the most massive galaxy detected in the observation with strong UV emission. The galaxy is centered at $\alpha =$ 14h08m13.59s, $\delta =$ 48d51m44.74s and z $=$ 0.052. The main aim of our present work is to dissect the properties of I~Zw~81 and understand the underlying physical mechanisms responsible for it's present state. While bright clusters and galaxy groups are expected to host galaxies of stellar masses up to $\log(M_\star/M_\odot)$ $\gtrsim$ 11.70, the void galaxies yet probed merely reach a stellar mass of $\log(M_\star/M_\odot)$ $=$ 11.20 \citep{SJPenny}. With stellar mass, $\log(M_\star/M_\odot)$ $\approx$ $10.90$ \citep[][]{2021Pandey}, I~Zw~81 stands amongst the massive galaxies detected in isolated environments. The galaxy has been previously reported in various observational surveys of the Bootes void \citep{1988seyfert,2002Cruzen}, and is known to host an Seyfert 2 AGN \citep{1988seyfert}. 

Few studies concentrating on the massive void galaxies (M$_\star$ $\gtrsim$ 10$^{10}$ M$_\odot$) reveal that such galaxies evolve passively contain old stellar population with low H$_\alpha$ SFRs ($<$ 1 M$_\odot$ yr$^{-1}$) and a handful of them host active nuclei \citep{SJPenny,2016mckelvie}. These results portray that regardless of the global environment, mass quenching mechanisms active in massive galaxies are capable of shutting down ongoing the star formation in a galaxy. However, the photometric properties of the galaxy in focus - I~Zw~81, show blue colors on the UV and optical color-magnitude diagrams along with strong far-UV (FUV) and near-UV (NUV) emission \citep{2021Pandey} which make the galaxy interesting to study.

Galaxies follow a strong morphology-density relation - `red and dead' galaxies with elliptical or S0 morphology are generally found in groups and clusters, whereas `star-forming' spirals or irregular galaxies are abundant in the field environment \citep{1980Dressler, 2010Peng}. At first, it appears that environment is the clear driver of evolution, but a few studies state that the stellar mass correlates better with the ongoing star formation and color than the local environment density \citep{2007Haines,2015Alpaslan}, though the results are debatable \citep{2004Balogh}. It is predicted that the optical emission profile of I~Zw~81 follows a De Vaucouleurs's light profile or has a large bulge surrounded by a faint disk \citep{1997Cruzen}. This structural description resonates with an early-type galaxy (ETG); thereby, I~Zw~81 seems to belong to a class of massive, blue ETGs situated in a void. While blue and low-mass ETGs are reported to exist in low-to-moderate density environment, there is a dearth of similar galaxies at the high stellar mass end (log($M_\star$/\msun) $\sim 11.2$). Most of the blue ETGs with stellar masses between 10$^{10.5}$ \msun and 10$^{11.2}$ \msun\ resemble major-merger remnants \citep[][]{2009kannappan}. Through this work, we investigate if the void galaxy has undergone a similar evolution scenario. 

The focus of our work is to understand the impact of mass quenching mechanisms in the absence of environmental quenching processes on the evolution of the blue and massive ETG.

\par
The paper is organized as follows: Section~\ref{sec:Data} describes the data used in this work. In Section~\ref{sec:LD}, we elaborate our structural analysis and discuss the derived properties of I~Zw~81. Section~\ref{sec:sed} summarizes SED fitting technique and derived results. Section~\ref{sec:centralSF} focuses on understanding the star formation activity in the galaxy. Finally, in Section~\ref{sec:dis}, we discuss and conclude our findings. In this work, we assume a flat cold dark matter cosmology model with $\Omega_m$ = 0.27, $\Omega_{\Lambda}$ = 0.73 and H$_0$ = 70 km s$^{-1}$ Mpc$^{-1}$. A unit arcsec is equivalent to 1 kpc at $z$ = 0.052.

% \begin{figure}
%      \centering
%      \includegraphics[width=0.7\linewidth]{figures/ds9.eps}
%      \caption{False color image of I Zw 81 - UVIT NUV (blue), $CFHT$ g (green) and $CFHT$ r (red). $CFHT$ images are convolved with PSF-matched kernels. The kernels were created on matching UVIT NUV with $CFHT$ g/r PSFs. }
%      \label{fig:rgb}
% \end{figure}

% \begin{table}
%     \centering
%     \caption{Basic information on I Zw 81}
%     \begin{tabular}{ccc}
%     \hline
%     \hline
%     Name & {\sc I} Zw 81  &  ref \\
%     \hline
%     RA (J2000) & 14 08 13.59   & a\\
%     DEC (J2000) & 48 51 44.74 & a \\
%     Distance & 230.7 Mpc & a\\
%     Scale (kpc/arcsec) & 1.012 & a\\
%     Galaxy type & Seyfert 2 & b \\
%     M$_{\rm B}$ (mag) & $-$20.3 & c \\
% %    Morphological type & S0b(r) & d \\
%     Inclination angle (deg) & 35.2 & c \\
%     \hline
%     \end{tabular} \\
%     {$^{\rm a}$ from NASA/IPAC Extragalactic Database (NED), $^{\rm b}$ \citet{1988seyfert}, $^{\rm c}$ \citet{1996Szomoru}}
%     \label{tab:IZw81}
% \end{table}

\section{Data and Observation}
\label{sec:Data}

% Table~\ref{tab:IZw81} summarizes the basic information on the galaxy.  While bright clusters and galaxy groups are expected to host galaxies of stellar masses up to $\log(M_\star/M_\odot)$ $\gtrsim$ 11.70, the void galaxies yet probed merely reach a stellar mass of $\log(M_\star/M_\odot)$ $=$ 11.20 \citep{SJPenny}. 
% With stellar mass, $\log(M_\star/M_\odot)$ $\approx$ $11$ \citep[this work,][]{2021Pandey}, I Zw 81 stand amongst the massive galaxies detected in isolated environments and is known to host an AGN \citep{1988seyfert,1992LINER}. 

I~Zw~81 was observed in our PILOT survey (Observing program G07-077, PI: Kanak Saha) of the Bootes void in FUV and NUV filters using UVIT onboard {\em AstroSat}. The observation was made simultaneously in UVIT F154W and N242W filters. The details of the observation and data reduction are mentioned in \citet{2021Pandey}. Complementary archival imaging data of the galaxy in optical and infrared wavebands were collected from the following surveys: $Sloan$ $Digital$ $Sky$ $Survey$ Data Release 12  \citep[$SDSS$ DR12,][]{2015Adr12}, 
$Canada$ $France$ $Hawaii$ $Telescope$ \citep[$CFHT$,][]{CFHT}, 
$Two$ $Micron$ $All$-$sky$ $survey$ \citep[$2MASS$,][]{2006Skrutskie} 
and $Wide$-$field$ $Infrared$ $Survey$ \citep[$WISE$,][]{2010AWise}. 
We have also collected mid-IR 60$\mu$m and 100$\mu$m fluxes of the galaxy from the
%observed with Infrared Astronomical Satellite \citep[IRAS,][]{1984Neugebauer}
catalogue of \citet{1990Moshir}.

\subsection{PSF matching and photometry}
\label{sec:phot}
Each imaging survey has a specific angular resolution. The full-width half maximum (FWHM) of the point spread function (PSF) corresponding to UVIT, $SDSS$ and $2MASS$ are $\sim$ 1\arcs.5, 1\arcs.2 and  2\arcs.8, respectively. Therefore, it is required to make the PSFs of each observation uniform prior to photometry. In our case, $2MASS$ has the poorest angular resolution amongst above mentioned observations. Hence, we use re-scaled $2MASS$ PSF to homogenize UVIT and $SDSS$ observations. We model $2MASS$ J-band PSF using the method described in Section~\ref{sec:skymask} which is further re-scaled to match the pixel scale of the targeted surveys (UVIT and $SDSS$). We create multiple PSF kernels combining re-scaled J-band PSF with PSFs of FUV, NUV and ugriz wavebands. The kernels were finally used to convolve UVIT and $SDSS$ image stamps of the galaxy. The entire process of PSF matching is carried out using Astropy {\tt PHOTUTILS} package {\citep{bradley_2020}}. 
\par
We perform fixed aperture photometry for all the ten bands (UVIT FUV/NUV, $SDSS$ ugriz, $2MASS$ J/H/Ks) using source extractor \citep[SExtractor;][]{1996Bertin}. The radius of the aperture chosen for the photometry is $\approx$ 10\arcs\ which is 2.5 times the optical effective radius of the galaxy (see Section~\ref{sec:structure}). Moreover, the surface brightness profile of the $2MASS$ J/Ks band reaches 20/ 21 mag arcsec$^{-2}$ at the same radius \citep{2006Skrutskie}. We also inspect the flux conservation of the source before and after convolution. The loss in flux due to convolution turns out to be much less than 1\% within a fixed aperture. The $WISE$ fluxes were measured within standard apertures having radius $\sim$ 8$^{\prime\prime}$.25 for W1 and W2 filters and 22$^{\prime\prime}$.0 for W3 and W4 filters \citep{2013wise}. The measured $WISE$ fluxes are aperture corrected prior to the use. We apply foreground extinction corrections to fluxes in all filters using the foreground IR dust maps from \citet{1998Schlegel}. Note that all the magnitudes/fluxes used are in AB magnitude system \citep{1983Oke}. The magnitudes/fluxes are further converted to mJy units prior to spectral energy distribution (SED) fitting. Wherever required, K-correction in magnitudes are done using publicly available codes given by \citet{2012Chiligan}.

\section{Light profile decomposition}
\label{sec:LD}
We use $CFHT$ Megacam publicly available deep imaging data for structural decomposition of the galaxy. The frames containing our galaxy are observed by $CFHT$ in g- and r-wavebands for $\sim$345s each. These frames are well-calibrated, reduced and astrometry corrected. The $CFHT$ observations of the galaxy (see left panel of Figure~\ref{tab:galfit}) suggests the presence of a ring along with a central component and a disk devoid of spiral arms.
 %\subsection{Sky subtraction and masking}

\subsection{Sky subtraction, masking and PSF modelling}
\label{sec:skymask}

We prepare cutout images centered around the galaxy from $CFHT$ frames of size $\approx$ 1.2$^{\prime}$ such that {$\sim$30\% of the cutout contains the sky background.} We run SExtractor over each $CFHT$ cutouts at a low detection threshold ($\approx$ 0.5) in order to generate the corresponding segmentation maps; setting a low detection threshold helps in identifying the faint features around the galaxy. These maps were used to mask the neighbouring sources around the galaxy. The estimated value of sky surface brightness in g- and r-bands obtained using SExtractor are $\approx$ 25.8 mag arcsec$^{-2}$ and 24.8 mag arcsec$^{-2}$, respectively. The sky background flux is subtracted from each cutout. The g- and r-band observation reaches a depth of 27.7 mag arcsec$^{-2}$ and 27.0 mag arcsec$^{-2}$, respectively.

Accurate modelling of PSF holds key importance in the process of structural decomposition. We select an isolated, bright and non-saturated point source from $CFHT$ g- and r-frames for PSF modelling. We use Moffat function for modelling as it models the wings of a PSF effectively compared to a Gaussian function \citep{2001Trujilo}. The functional form of the Moffat profile used is given by:

\begin{equation}
 f(x, y) = f_0 \left[1 + \frac{ (x - x_0)^2 + (y - y_0)^2 } { \alpha^2 }  \right]^{-\beta}
\label{eq:Moffat}
\end{equation}
 
\noindent with FWHM = $2\alpha \sqrt{2^\frac{1}{\beta}-1}$. Here, $f_{0}$ is the central flux and $\beta$ and $\alpha$ are the free parameters in the function. We model the selected point source using 2D Moffat function present in the Astropy package. The resultant PSF FWHM in the $CFHT$ g- and r-band are  0\arcs.6 and 0\arcs.74, respectively.

\subsection{Structural Analysis}
\label{sec:structure}
GALFIT is commonly used to model the light profile of a galaxy using various ellipsoidal models \citep{2002Peng}. We use GALFIT for 2D modelling of $CFHT$ g- and r-band images of {\sc I} Zw 81. The required weight ($\sigma$) images were generated internally using the value of GAIN, RDNOISE and NCOMBINE present in the header of our input images by following Poisson statistics. We use a combination of the following analytical profiles in our multi-component structural modelling.

\par
i). {\bf Sersic function:} The function is one the most used function to study galaxy morphology. It is defined as follows:

\begin{equation}
\Sigma(r) = \Sigma_e \exp \Bigg(-\kappa \bigg[\Big(\frac{r}{r_e} \Big)^\frac{1}{n}-1 \bigg] \Bigg) 
\label{eq:sersic}
\end{equation}

\noindent where $\rm \Sigma_e$ is the surface brightness at the effective radius r$\rm _e$. Here, {$n$ is the Sersic index whose value increases as profile of the component gets steeper in the inner part and flatter in the outskirt.} The factor $\kappa$ is the normalization constant dependent on $n$. The parameters obtained in the fit are: center coordinates, integrated magnitude (m$_0$), r$_e$, $n$, axis ratio (q) and position angle ($\theta_{\rm PA}$). 

ii). {\bf Exponential disk function:} The exponential function (expdisk) is described as follows:
\begin{equation}
    \Sigma (r) = \Sigma_{0} \exp \Big(- \frac{r}{r_s} \Big) 
\end{equation}
\noindent where $\Sigma_{0}$ is the central surface brightness of the disk and r$_{s}$ is the scale length of the disk. Here, the output parameters are: central coordinates, m$_0$, r$_s$, q and $\theta_{\rm PA}$.

\par
We run GALFIT over g-band image with a single Sersic function which gives a resultant $n$ $=$ 5.07 and r$\rm _e$ $\approx$ 4$^{\prime\prime}$.16. In addition, we also obtain the central coordinates of the galaxy (x$_0$, y$_0$) from the fit. {An exponential function is added in the input to fit the disk while existing Sersic function models the central component. The resultant output obtained by combining the two functions improves the fit but leaves a significant amount of positive fluxes in the residue. Thereafter, we added another Sersic component and expdisk function to model the bar and outer disk, respectively. {An additional parameter $c$ is appended with the Sersic function to model the bar to find whether the bar is disky or boxy in shape. The r$_e$ corresponding to the Sersic function used to model the central component successively came out to be less than the PSF FWHM which implies that the central component is small and unresolvable.} Hence, we replaced the central Sersic component with a PSF to model the component}. The central coordinates of both the inner and outer disks were fixed to (x$_0$, y$_0$) while all other parameters of different components were left free throughout the run. We also model the sky background for each runs simultaneously.
\par

iii). {\bf Truncated exponential function}: We venture to model the ring using a truncated exponential disk function \citep{2010PengGF}. A truncated exponential disk function in GALFIT is primarily a hyperbolic tangential function whose functional form is as follows:

\begin{equation}
    P (x, y) = \tanh{\rm (x, y; x_0, y_0, r_{break}, \Delta r_{\rm soft}, q, \theta_{PA} ) }
\end{equation}

\noindent where (x$_0$, y$_0$), $q$ and $\theta_{\rm PA}$ are the central coordinates, axis ratio and position angle of the ring. The parameter $r_{\rm break}$ is the break radius at which the truncated model flux is 99\% of the original un-truncated model flux at that radius while $\Delta r_{\rm soft}$ is the softening length where r$_{\rm break}$ $\pm$ $\Delta r_{\rm soft}$ is the radius where the truncated model flux drops to 1\% of its original flux at the same radius.  The inner and outer part of the exponential disk function is modified with the same radial truncation function in our case. The free parameters in the function are: central coordinates, r$_{\rm break}$, surface brightness at r$_{\rm break}$, r$_s$, q, $\Delta r_{\rm soft}$, $\theta_{\rm PA}$, etc. 

To account for the observed lopsidedness of the ring, we introduce Fourier modes to the radially truncated function in order to match the actual form of the ring. The Fourier modes perturbs a perfect ellipsoid as follows:

\begin{equation}
    r(x, y) = r_0 (x, y)\Big(1 + \sum^{N}_{m = 1} a_m \cos{m(\theta +\phi_{m})}\Big) 
\end{equation}

\noindent where $r_0$(x, y) is the unperturbed radial coordinates, $a_{\rm m}$ is the Fourier amplitude for mode $m$, $\theta$ $=$ $\arctan{((y - y_0)/(x - x_0)q)}$, $N$ is the total number of modes specified by the user and $\phi_m$ is the phase angle of mode $m$ relative to $\theta$. 

\begin{figure*}
    \centering
    \includegraphics[width=0.80\linewidth]{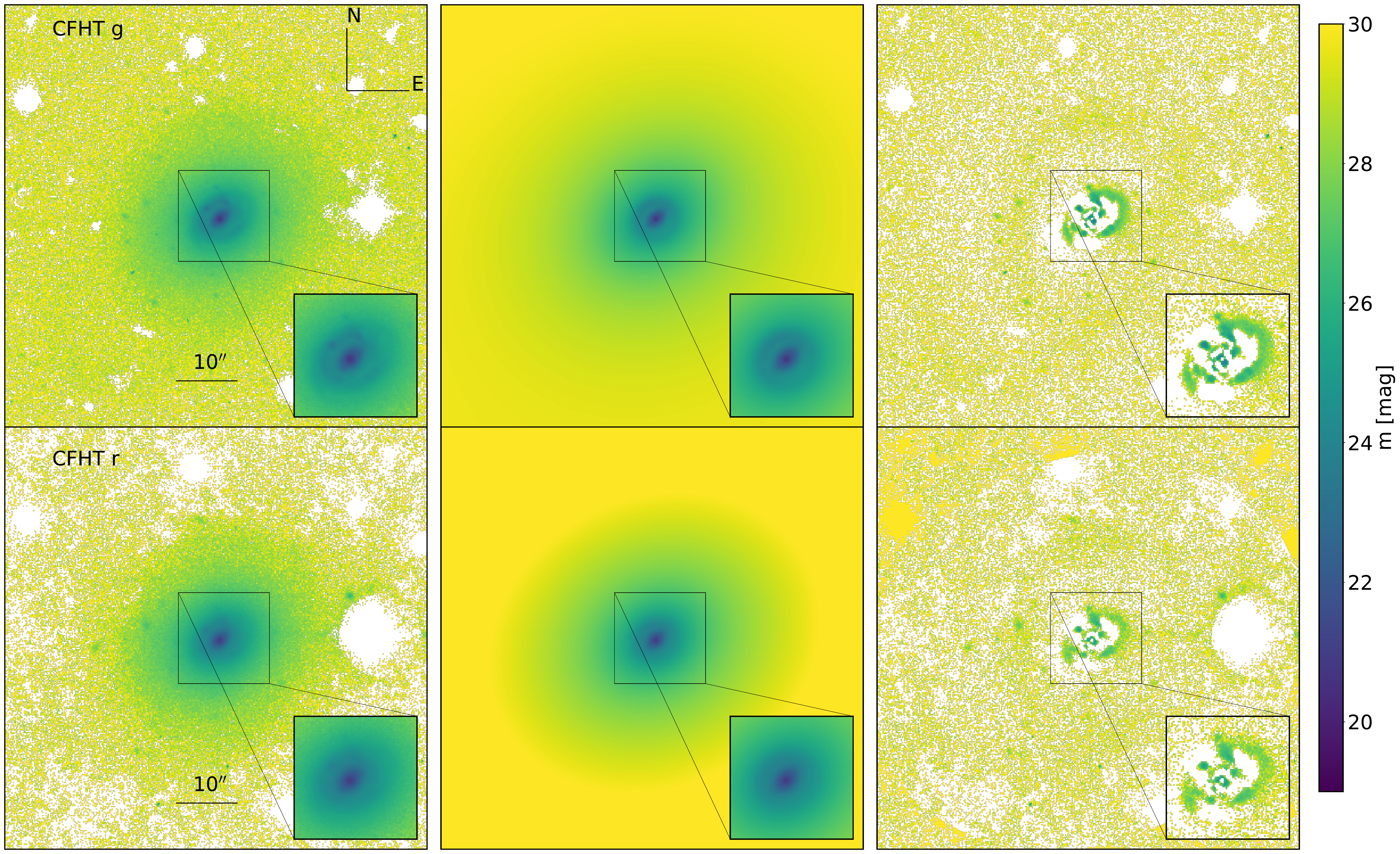}
    \includegraphics[width=0.7\linewidth]{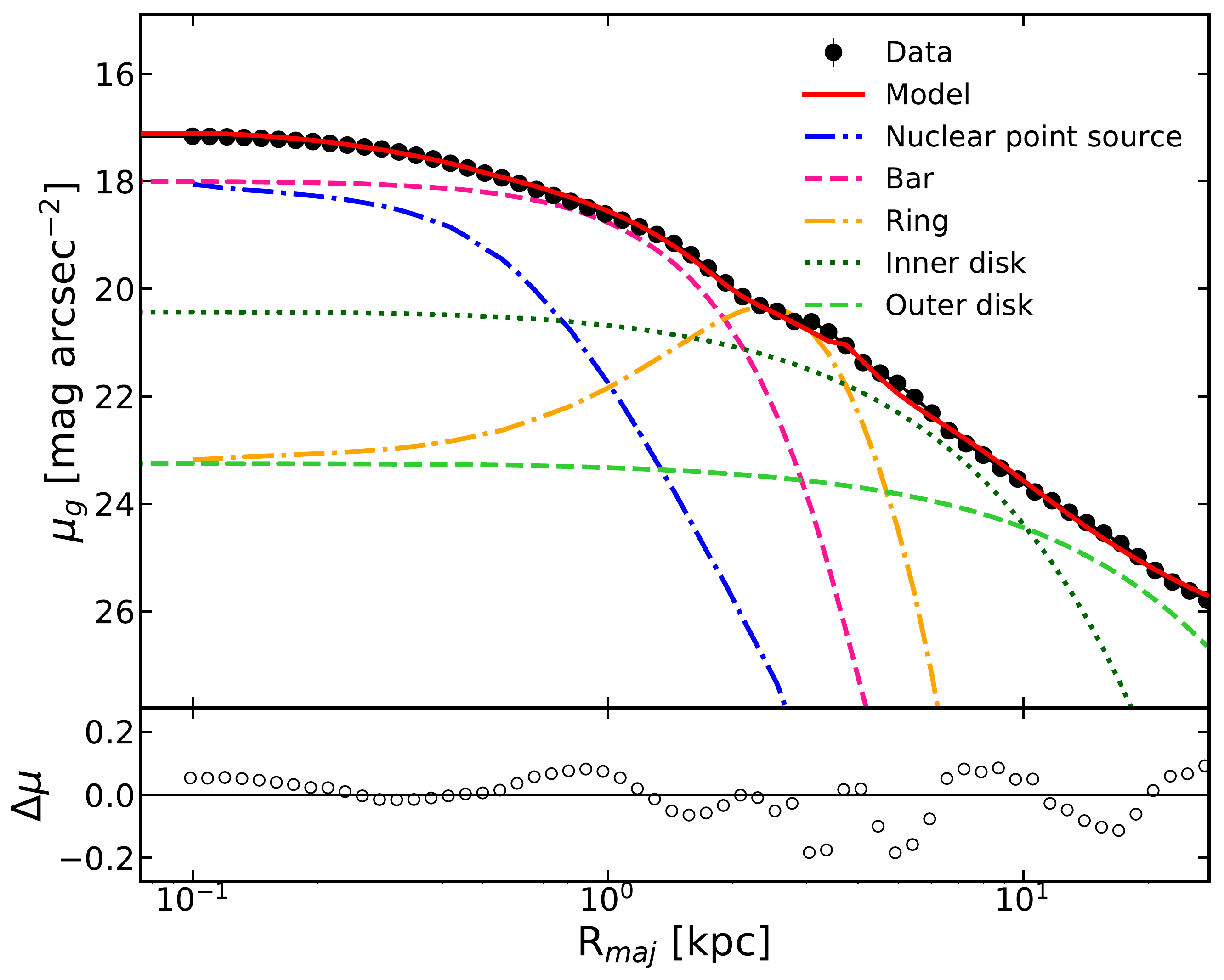}
    \caption{{\it Upper panel}: From left to right in upper ($CFHT$ g-band) and lower ($CFHT$ r-band) segments: Image of I~Zw~81, GALFIT best-fit model image, and corresponding residual image in magnitude units. The inset images in all images highlight the inner region of the galaxy. {{\it Bottom panel}: Isophotal surface brightness profile of the galaxy in $CFHT$ g-band. Solid black circles and red line show the observed and best-fit model profiles, respectively. The surface brightness distribution of the individual component is coded with different line styles and colors, as mentioned in the legends. The residual obtained by subtracting model surface brightness from that of the galaxy is shown in the lower segment of the bottom panel.}}
    \label{fig:galfit_output}
\end{figure*}

We conduct subsequent re-runs combining all the analytical functions discussed above, i.e., a PSF, Sersic, two expdisk and truncated exponential functions, until the $\chi_{\nu}^{2}$ is minimized and the residual image look random and balanced with both negative and positive flux values. {The fraction of total light recovered by our best-fit models is close to unity in each band.} The $\chi_{\nu}^{2}$ values for our final fitting of $CFHT$ g- and r-bands are 2.12 and 1.03, respectively. The upper panel of Figure~\ref{fig:galfit_output} shows the $CFHT$ g- and r-band observed images of the galaxy, best fit models and their corresponding residuals. {The lower panel of the figure shows 1D surface brightness profiles of the galaxy, model and its constituent components. The isophotal fittings of the model, components and the galaxy were performed using PHOTUTILIS.} Table~\ref{tab:galfit} gives the details of our GALFIT output parameters for the $CFHT$ g- and r-band models. We calculated the light fraction for each component of the galaxy in g-band using m$_0$ given by GALFIT. The total emission from the ring is obtained by integrating its flux contribution from the origin to r$_{\rm break}$. The resultant values of light fraction for each component are mentioned in Table~\ref{tab:galfit}. {The choices of the chosen components for fitting are explained in the following section.} 

%KS: done

\par

\subsection{Interpretation of model parameters}
\label{sec:interpret}
The overall light profile of the galaxy is disky with no hint of spiral arms. Hence, we categorize I~Zw~81 as a lenticular galaxy. We exploit the GALFIT output parameters to briefly discuss the structural and photometric properties of all components of galaxies.

\subsubsection{{Nuclear point source}, bar and ring}
Bulges classify into two categories, namely, classical bulges and pseudo-bulges \citep{2008Fisher,kormendy2016}. The Sersic index, $n$ is a robustly used tool to differentiate bulges such that $n$ $>$ 2 for classical bulges and $n \leq 2$ in the case of pseudo-bulges. However, as the central component of I~Zw~81 is modeled with a PSF function, we do not have the values of $n$.{The fraction of light enclosed by the component is $\approx 0.06$.

The bar-driven secular evolution transforms a small round classical bulge by flattening it structurally and changing it kinematically to appear like a boxy-pseudobulge \citep{2012Sahaetal,2015Saha}. These authors have shown that a small classical bulge (bulge-to-total light ratio $\sim 0.06-0.08$) in Milky Way-like barred galaxies could remain hidden. I Zw 81, being barred and more massive than the Milky Way, and having nuclear component light fraction $\sim 0.06$, certainly holds the possibility of a hidden small classical bulge. \citet{2017Neumann} also indicated the presence of a nuclear disk in secularly evolving galaxies. The galaxy also hosts an AGN, which may partly contribute to the observed light fraction. The overall structure of the central component remains unclear. We refer to the central component as a nuclear point source (NPS).}

%It was shown by \citet{2012Sahaetal} that the bar-driven secular evolution transforms low-mass classical bulges by flattening them kinematically, acquiring spheroidal shape. \citet{2015Saha} worked with galaxies less massive than I Zw 81 having comparable central component light fraction. The author demonstrated that the photometric detection of small classical bulges in barred galaxies is difficult. Secularly evolving galaxies show signatures of small nuclear disks \citep{2017Neumann}, which is unresolvable with CFHT observations. The galaxy hosts an AGN that may contribute to the observed light fraction. The overall structure of the central component is unclear. We refer to the component as a nuclear point source (NPS).}

%\citet{Gao_2020} argued that the parameter bulge-to-total light ratio is not a good discriminator of bulge type. 

\begin{table*}[h]

    \centering
    \caption{GALFIT 2D best fit results for the g- and r-bands.}
    \begin{tabular}{cccccccccccc}
    \hline
    \hline
    Components & \multicolumn{1}{c}{{NPS}} & \multicolumn{4}{c}{Bar} & \multicolumn{1}{c}{Ring}
    & \multicolumn{2}{c}{Inner disk (ID)} & \multicolumn{2}{c}{Outer disk (OD)} \\
    \hline
    Galfit function & \multicolumn{1}{c}{PSF} & \multicolumn{4}{c}{Sersic} &
    \multicolumn{1}{c}{Trunc. expdisk}&
    \multicolumn{2}{c}{expdisk} & \multicolumn{2}{c}{expdisk} \\
    \hline
    Band & $\rm (m_{0})_{NPS}$ & (m$_{\rm0})_{\rm bar}$ & $\rm (r_e)_{\rm bar}$ & n$_{\rm bar}$ & c$_{\rm bar}$ & r$_{\rm break}$ & (m${\rm _0})_{\rm ID}$ & $\rm (r_s)_{\rm ID}$ & (m${\rm_0)}_{\rm OD}$ & $\rm (r_s)_{\rm OD}$ \\
     & (mag) & (mag) & (arcsec) & & & (arcsec) & (mag) & (arcsec) & (mag) & (arcsec) \\
    (1) & (2)& (3)& (4)& (5)& (6)& (7)& (8) & (9) & (10) & (11)  \\
    \hline
    g & 18.52 & 16.90 & 0.92 & 0.57 & $-$0.32 & 4.20 & 16.36 & 2.04 & 16.53 & 7.35\\
    r & 17.84 & 16.42 & 0.92 & 0.65 & $-$0.31 & 4.59 & 15.74 & 1.93 & 15.80 & 7.57\\
    \hline
    Light fraction (g-band) & \multicolumn{1}{c}{0.06} & \multicolumn{4}{c}{0.23} &
    \multicolumn{1}{c}{$<$0.1} &
    \multicolumn{2}{c}{0.38} & \multicolumn{2}{c}{0.32} \\
    \hline
    \end{tabular}
     
     { Col. (2): Integrated magnitude of the {nuclear point source (NPS)}. Col. (3)-(6): Integrated magnitude, effective radius, Sersic index and boxy/disky parameter of the bar. Col. (7): break radius of the ring. Col. (8) - (9): Integrated magnitude and disk scale length of inner disk. Col. (10) - (11): Integrated magnitude and disk scale length of outer disk. The last row represents the light fraction of all components of in the $CFHT$ g-band, respectively. The magnitudes in the table are not corrected for Galactic extinction. 
     }
    \label{tab:galfit}
\end{table*}

I~Zw~81 hosts a {disky} bar with r$_e =$ 0.921 kpc, $n$ $\approx$ 0.6 and $c < 0$. The flat intensity profile of the bar resembles those typically seen in ETGs \citep[][]{flatbar}. The bar-to-total light ratio (Bar/T) is estimated to be 0.23, suggesting the presence of a strong bar \citep{2020Barway}. Note that the incidence of a strong bar is not considered so common in S0 galaxies \citep[][]{2010Buta}. The stellar ring is placed close to the bar-end, possibly at the bar corotation resonance. The outer radius of the ring (r$_{\rm break}$) is 4.2 kpc. The light emitted by the ring amounts to less than 1\% of the total light inside the galaxy.

\subsubsection{Inner and outer disks}
The structural decomposition of the galactic disk in I~Zw~81 reveals a `break' in its radial profile, {wherein the break radius is $\approx$10 kpc (see lower panel of Figure~\ref{fig:galfit_output}).} The galaxy is disk-dominated as both of its disks contribute to $\approx$70\% of the total light in the galaxy (see Table~\ref{tab:galfit}). Using the integrated magnitudes in $CFHT$ g- and r-bands given by GALFIT (Table~\ref{tab:galfit}), we estimate the central surface brightness of the inner disk (ID) and outer disk (OD) using the following equation~\ref{eq:SB0}:

\begin{equation}
    \mu_0 = m_0 + 2.5\log(2 \pi r_s^2)
    \label{eq:SB0}
\end{equation}

The central surface brightness, $\mu_{\rm 0,ID}$ and $\mu_{\rm 0,OD}$ are further corrected for inclination and cosmological dimming effects using the following equations \citep{1930PTolman,2008Zhong}:

\begin{equation}
    \mu_{0,c} = \mu_0 + 2.5\log(q) - 10\log(1+z).
    \label{eq:cSB0}
\end{equation}

\noindent Here, $q$ is the axis ratio of the disk and $z$ is the redshift of the target galaxy. We proceed to determine the central surface brightness of the disks in B-band using the following relation \citep{2002Smith}:

\begin{equation}
    \mu_0(B) = \mu_{0,c} (g) + 0.47(\mu_{0,c} (g) - \mu_{0,c} (r)) + 0.17
    \label{eq:BSB}
\end{equation}

\noindent Using equations~\ref{eq:SB0},~\ref{eq:cSB0} and~\ref{eq:BSB}, the central surface brightness of the inner disk, $\mu_{\rm 0,ID} \rm (B)$ and outer disk, $\mu_{\rm 0,OD} \rm (B)$ obtained are 20.31 mag arcsec$^{-2}$ and 23.32 mag arcsec$^{-2}$, respectively. The central surface brightness, $\rm \mu_{B}$ $\geq$ 22.5 mag arcsec$^{-2}$, separates low surface brightness (LSB) and high surface brightness regime \citep[HSB;][]{2009Rosenbaum}. We conclude that the inner region or the inner disk is at HSB surrounded by a LSB outer disk.

Our structural analysis reveals that the galaxy is disk-dominated with a well-defined bar and has no apparent sign of a classical bulge that could have formed due to major mergers \citep{kormendy2016}. Based on a suite of GALMER simulations, it is shown that in no case a major-merger S0-like remnant form a significant stellar bar \citep{2018moral}. It is likely that I~Zw~81 being in the void region, managed to avoid any major-merger-like events during its history and evolved secularly \citep{KormendyKennicutt2004}.

% \begin{table}
%     \centering
%      \caption{Details of wavebands used during SED fitting.}
%     \begin{tabular}{ccc}
%     \hline 
%     \hline
%   Filter & $\lambda (\angstrom)$ & AB Magnitudes (mag)\\
%     \hline
%     UVIT FUV & 1541 & 19.07 \\
%     UVIT NUV & 2418 & 18.39 \\
%     $SDSS$ u & 3560 & 17.05\\
%     $SDSS$ g & 4680 & 15.65 \\
%     $SDSS$ r & 6180 & 14.97 \\
%     $SDSS$ i & 7500 & 14.59 \\
%     $SDSS$ z & 8870 & 14.36 \\
%     $2MASS$ J & 12400 & 13.03 \\
%     $2MASS$ H & 16600 & 13.82 \\
%     $2MASS$ Ks & 21600 & 13.80 \\
%     $WISE$ W1 & 34000 & 14.35\\
%     $WISE$ W2 & 46000 & 14.71 \\
%     $WISE$ W3 & 120000 & 12.93 \\
%     $WISE$ W4 & 220000 & 12.31 \\
%     IRAS 60$\mu$m & 600000 & 2.40 (0.395 Jy)$^*$\\
%     IRAS 100$\mu$m & 1000000 & 1.39 (1.00 Jy)$^*$\\
%     \hline
%     \end{tabular}\\
%   %  {}
%     \label{tab:data}
% \end{table}

\section{SED modelling and physical parameters}
\label{sec:sed}
%W1 ($\lambda_{\rm eff}$ = 3.4${\mu}$m), W2 ($\lambda_{\rm eff}$ = 4.6${\mu}$m), W3 ($\lambda_{\rm eff}$ = 12${\mu}$m) and W4 ($\lambda_{\rm eff}$ = 22${\mu}$m) wavebands

We use CIGALE (Code Investigating GALaxy Emission) for SED fitting of the galaxy to derive its physical properties such as stellar mass, SFR, dust attenuation, etc. \citep{2019Boquien}. CIGALE is a python based approach to model UV to IR emission of a galaxy. It uses the Bayesian inference paradigm to determine the model parameters. The 16 waveband fluxes used by us for SED fitting are, UVIT FUV/NUV, {\em SDSS} u/g/r/i/z, {\em 2MASS} J/H/Ks, {\em WISE} W1/W2/W3/W4, and {\em IRAS} 60$\mu$m/100$\mu$m. The procedure used for the flux measurement in respective bands is described in section~\ref{sec:phot}. CIGALE offers multiple modules to model the star formation history (SFH), dust attenuation, and stellar population of a galaxy. The modules used to model each physical process are given below: \\

\indent{(i). We select {\tt sfh2exp} module to predict the SFH of the galaxy. The model comprises of two exponentially decaying functions where the first function is used to model the long term stellar population and the second function models the most recent starburst. Thus, the SFH model is governed by five parameters, i.e., the age of the main stellar population in the galaxy, the e-folding time of the main stellar population ($\tau_{\rm main}$), the e-folding time of the late starburst population ($\tau_{\rm burst}$), the age of the late burst population (age$_{\rm burst}$), and the mass fraction of the late burst population (f$_{\rm burst}$).}

\indent{(ii). We adopt \citet{2003Bruzual} library of single stellar population to model the stellar emission of our galaxy with \citet{1955Salpeter} initial mass function keeping metallicity fixed at 0.02. We use N2 method to calculate the oxygen abundance of the galaxy \citep{2004Pettini}.}% section~\ref{sec:met}. }

\indent{(iii). We include {\tt nebular} module to estimate the nebular emission from our galaxy. The ionization parameter (log U) is fixed at $-2.5$, fraction of lyman continuum photons escape (f$_{\rm esc}$) and fraction of lyman continuum photons absorbed (f$_{\rm dust}$) varies between 0 and 0.1.}

\indent{(iv). To account for the dust attenuation caused to the stellar and nebular emission, we employ {\tt dustatt\_modified\_starburst} module \citep{2000Calzetti}. The value of internal reddening for nebular emission, $\rm E(B-V)_{nebular}$, ranges between 0.37 and 0.45 in steps of 0.4. The aforementioned range encloses $\rm E(B-V)_{nebular}$ ($=$ 0.41) computed using Balmar decrement method \citep{2006Osterbrock}. The reduction factor applied to the old stellar population varies between 0.4 and 0.6 in steps of 0.1. The attenuation curve is multiplied by a power-law in wavelength, $\lambda^{\delta}$, where $\delta$ varies between $-0.1$ and $-0.5$, and the amplitude of the bump at 2175 \AA\ fixed at $3.0$}. 

\indent{(v). %The total luminosity absorbed is re-radiated at mid and far-IR wavelengths due to the presence of polycyclic aromatic hydrocarbon, and dust grains.
The dust emission process is modelled adopting {\tt Dale 2014} module \citep{2014Dale}. The star-forming component in a galaxy is parameterized by a single parameter $\alpha$ defined as: $\rm d M_d (U) \, \propto U^{-\alpha} \, dU$, where M$_d$ being the dust mass, and $U$ the radiation field intensity. Here, $\alpha$ is fixed at $-2.0$. We include $f_{\rm AGN}$ to separate AGN contribution to the infrared dust emission.} 

\indent{(vi). We adopt {\tt fritz2006} AGN template in our SED fitting to model AGN emission. The module deals with seven free parameters: r is the ratio of maximum to minimum radii of the dust torus, $\beta$ and $\gamma$ describe the dust density with respect to the radius and opening angle ($\theta$), $\psi$ is the angle between AGN axis and the line of sight and {fracAGN is the ratio between AGN IR luminosity to total IR luminosity.}}

\begin{figure}
    \centering
    \includegraphics[width = 0.99\linewidth]{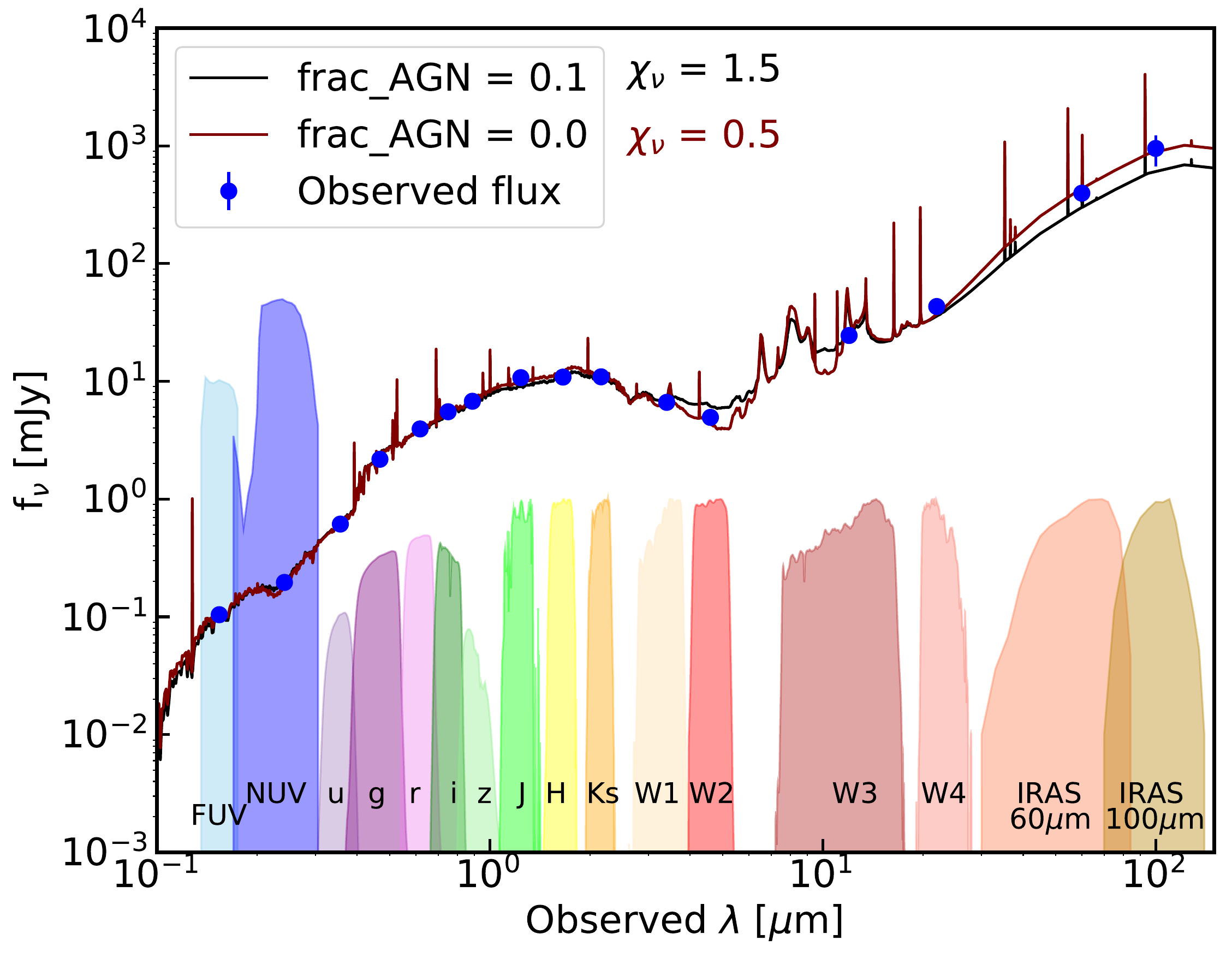}
    \caption{Best-fit SEDs (fracAGN $= 0./0.1$) of I~Zw~81 produced by CIGALE. The input broadband fluxes are marked with blue circles. The best fit parameters can be found in Table~\ref{tab:sed}. The error bars on the measured fluxes denote $1\sigma$ error bar. The filter response curves are shown on the figure with different colors. %The reduced $\chi^2_\nu$ value corresponding to both best-fit models are written on the figure.
    }
    \label{fig:sed}
\end{figure}

Table~\ref{tab:sed} comprises of all the selected modules and input parameters. The best-fit parameters are bold-faced. The combined grid parameters yield 139,96,800 model templates. The best-fit SED model is shown in Figure~\ref{fig:sed}.

{The best-fit value of fracAGN is $0.0$, which implies that the obtained SED is compatible with no AGN contribution to the total IR emission. We set the value of fracAGN to 0.1 (next admissible value) and compute the model SED and various physical parameters. The values of $\chi^2_\nu$ at fracAGN = 0.0 and 0.1 are 0.5 and 1.5, respectively. We show the best-fit model SEDs corresponding fracAGN = 0./0.1 in Figure~\ref{fig:sed}. The physical properties of I Zw 81 show slight variation in either case, suggesting that both values are equally likely to model the broadband emission of the galaxy. However, the best-fit value of the parameter never exceeds 0.1. Hence, we suspect the presence of a low luminosity AGN \citep{2018Ciesla}.}

%The derived physical quantities of the galaxy show slight variation in both cases implying that both values could be equally likely. However, we noticed that the best-fit values of the parameter never exceeded 0.1. Hence, we suspect there might be presence of a low luminosity AGN \citep{2018Ciesla}.}

The stellar mass of the galaxy obtained from our SED fitting is $\sim 7.8 \times 10^{10} M_\odot$. 
The age of the main stellar population is 8 Gyr, and the mass ratio of young-to-old stars is $\approx$ 0.001, implying the presence of a dominant fraction of old and evolved stars in the galaxy. We compare the properties of I~Zw~81 with a sample of bright S0 galaxies from \citet{2013Barway}. The stellar population age of the bright S0 galaxies is more than 9 Gyr, whereas D$_{\rm n}$(4000) index \citep{1999Balogh} for the sample peaks at 2.0 and {\em GALEX} NUV$-$r color for such galaxies is dominantly greater than 5.4 mag. These galaxies are generally present in groups and clusters. In the case of I~Zw~81, the NUV $-$ r color is 3.35 mag ({\em GALEX} NUV $-$ r = 3.46 mag) and D$_{\rm n}$(4000) = 1.57. These parameters suggest that I~Zw~81 is more star-forming, and the stellar population age of the galaxy is lower than the bright S0 galaxies studied by \citet{2013Barway}. 

\begin{figure*}
    \centering
    \includegraphics[width=0.99\linewidth]{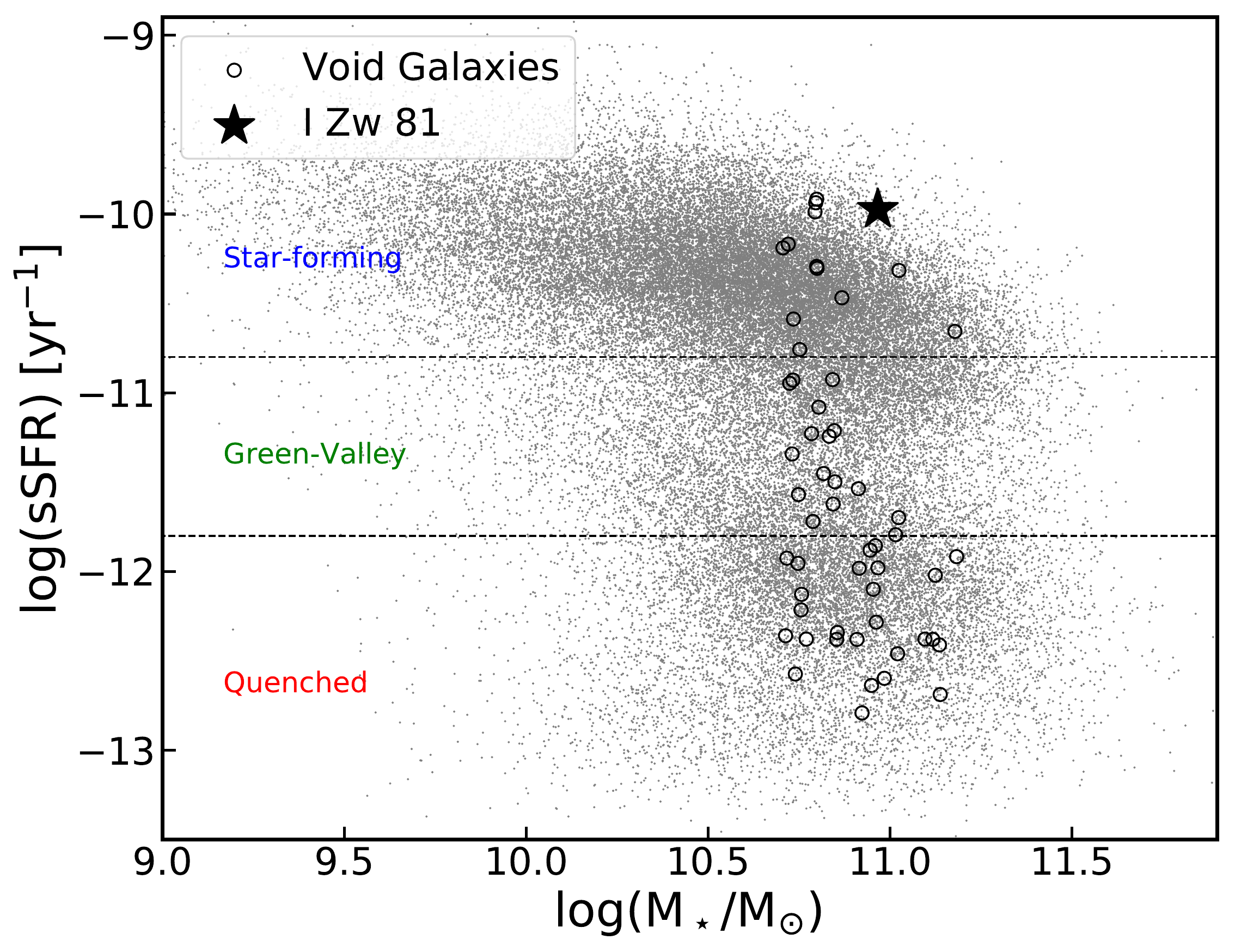}
    \caption{M$_\star$ versus sSFR distribution for void galaxies. Black star marker denotes IZw81.  Black open circles are void galaxies with M$_\star$ $>$ 5 $\times$ 10$^{10}$ \msun\ \citep{pan2012}. The background galaxies (grey dots) are taken from \citet{2016Salim}. The figure is divided into three sections: star-forming region ($\log\, sSFR\, \geq\, -10.8$), green valley region ($-10.8 > \log\, sSFR\, >\, -11.8$), and quenched region  \citep[$\log\, sSFR\, \leq\, -11.8$;][]{2020Barway}.}
        \label{fig:ssfr}
\end{figure*}

Using SFR and M$_\star$ obtained from SED fitting, we construct the sSFR vs. M$_\star$ diagram for the galaxy as shown in Figure~\ref{fig:ssfr}. The background galaxies, showing bimodal distribution are taken from \citet{2016Salim}. The void galaxies with M$_\star$ $>$ 5 $\times$ 10$^{10}$ \msun\ present in the figure are from \citet{pan2012}. We procure the values of SED-generated sSFRs and M$_\star$ of these void galaxies from Galaxy and Mass Assembly DR2 catalog \citep{2015gama}. Figure~\ref{fig:ssfr} shows that a majority of massive void galaxies evolve passively while a small fraction of them, including I~Zw~81 lie in the star-forming region. Our analysis confirms that the massive I~Zw~81 is star-forming and hosts a younger stellar population than its counterparts in groups and clusters.

\begin{table}
 \caption{CIGALE SED fitting parameters. The best fit parameters are bold-faced. }
    \centering
    \begin{tabular}{cc}
    \hline
    \hline
    Parameters & Values \\
    \hline
    \multicolumn{2}{c}{Star formation history - {\tt sfh2exp}} \\
    \hline
    $\rm \tau_{main}$ (Myr)  &  3000, 5000, 7000, \textbf{9000}, 10000 \\
    $\rm age$ (Myr) & \textbf{8000}, 9000, 10000, 11000, 12000  \\
    $\rm \tau_{burst}$ (Myr)  & 50 , \textbf{100}, 500, 1000, 2500, 5000 \\
    $\rm burst_{age}$ (Myr)   &   \textbf{500}, 1000, 1500, 2000\\
    $\rm f_{burst}$   & 0.01, \textbf{0.1}, 0.25, 0.50  \\
     \hline 
    \multicolumn{2}{c}{\tt nebular}  \\ 
    \hline
    f$_{\rm esc}$ & \textbf{0.0}, 0.1 \\
    f$_{\rm dust}$ & \textbf{0.0}, 0.1\\ 
    \hline
    %  & Line width (km/s) & 300 \\
    % \hline
    \multicolumn{2}{c}{Dust attenuation - {\tt dustatt\_modified\_starburst}} \\
    \hline
     E(B$-$V)$_{\rm nebular}$ & 0.37, \textbf{0.41}, 0.45 \\
    E(B$-$V)$_{\rm factor}$  & 0.4, \textbf{0.5}, 0.6 \\
    powerlaw\_slope & $-0.1$, $\bf -0.3$, $-0.5$  \\
    \hline
    \multicolumn{2}{c}{Dust emission - {\tt Dale 2014}}\\
    \hline
    $f_{\rm AGN}$ & \textbf{0.0}, 0.1 \\
    % Power law slope ($\alpha$) &  \textbf{2.00} \\
    \hline
   \multicolumn{2}{c}{AGN - {\tt Fritz 2006}}\\
    \hline
    r\_ratio & 30,60,\textbf{100}\\
%    $\tau$ & 1.\\
    $\beta$ & $-$0.75, \textbf{$-$0.25} \\
    $\gamma$ & 2.0, \textbf{4.0}, 6.0 \\
    opening\_angle & 60, 100, \textbf{140}\\
    psy & 30.1, \textbf{70.1}\\
    fracAGN & \textbf{0.0}, 0.1, 0.2\\
    \hline
    \end{tabular}
   
    \label{tab:sed}
\end{table}

\section{Central star-formation and the blue bar}
\label{sec:centralSF}

\begin{figure*}
    \centering
    \includegraphics[width=1.0\linewidth]{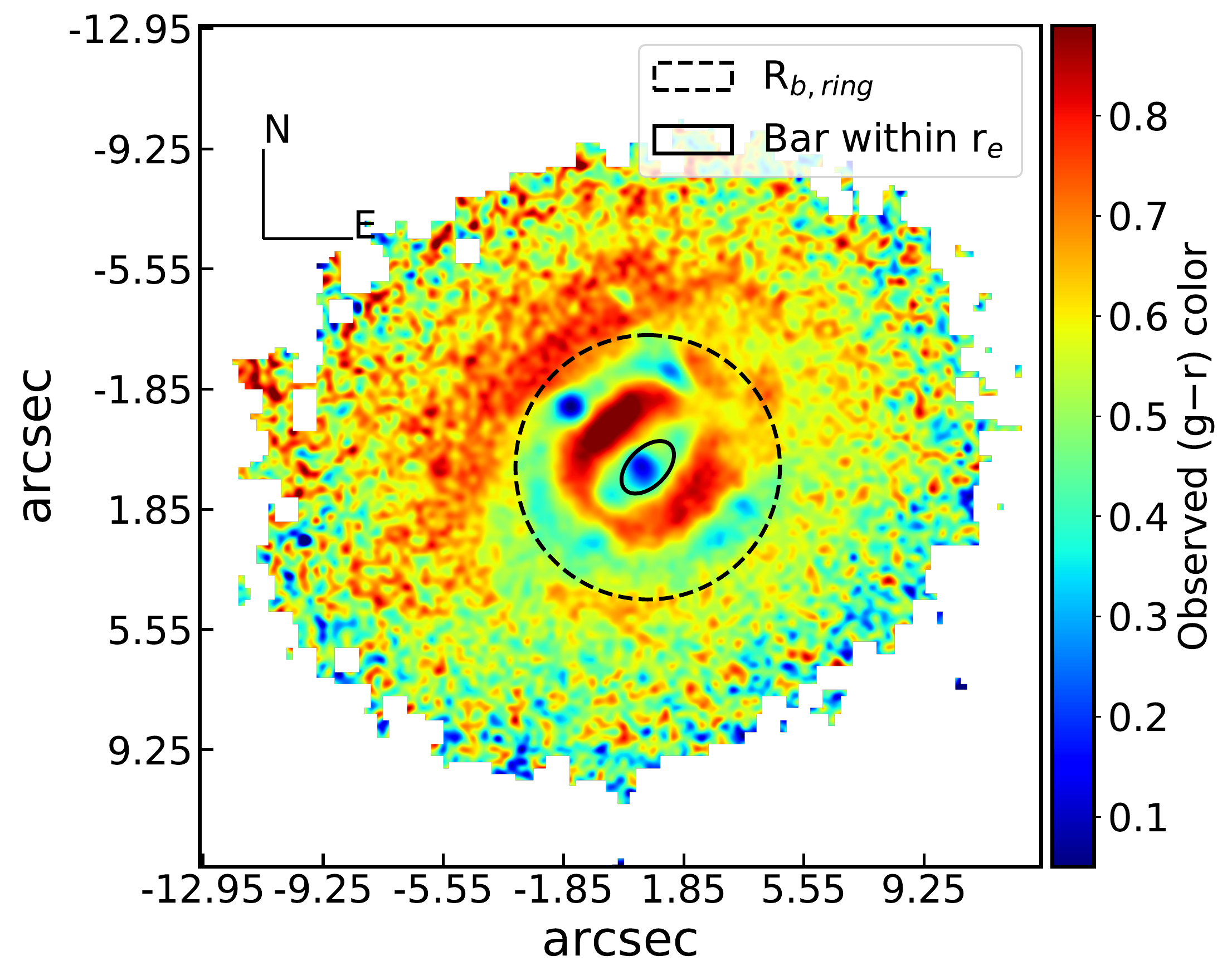}
    \caption{$CFHT$ g$-$r color map of I~Zw~81. Colors shown are not corrected for Galactic or internal extinction. Black solid ellipse traces the GALFIT modelled bar (r$_{\rm e}$ = 0.92\arcs, $\theta_{\rm P.A.}$ =  45$^\circ$, q = 0.55) whereas the radius of black dotted circle denote r$_{\rm break}$(= 4.2\arcs).
    }
    \label{fig:gr_cm}
\end{figure*}

Figure~\ref{fig:ssfr} shows that the sSFR of I~Zw~81 lie nearly half a dex above the star-forming main sequence. We inspect if such vigorous star formation activity is global or localized to certain regions in the galaxy with the help of $CFHT$ observation and GALFIT model parameters. Figure~\ref{fig:gr_cm} depicts the (g$-$r) color map of I~Zw~81 wherein the outer radii of the stellar ring and the bar within r$_e$ are shown. The bar appears to be optically blue while there is an asymmetry in the color of the disk inclining towards the redder side of the color scale.  

We use the integrated magnitudes provided in Table~\ref{tab:galfit} to find colors of individual components and qualitatively consider them as a proxy for the stellar population age. The observed optical colors (g$-$r)$_{\rm bar}$, (g$-$r)$_{\rm ID}$ and (g$-$r)$_{\rm OD}$ are 0.44 mag, 0.58 mag, and 0.69 mag, respectively. It turns out that the bar is the bluest, and the outer LSB disk is the reddest in our galaxy like the case of Malin~1 whose central region has an S0 morphology with a bluer bar \citep{Sahaetal2021}.
%This is in contradiction with the negative color gradient commonly seen in early-type S0 and elliptical galaxies \citep{1976Strom,2018Marian} but see \cite{Sahaetal2021} for the blue bar in Malin~1 hosting S0-like morphology in its central region. 
Based on optical colors, we argue that I~Zw~81 contains a mix of old and young stars, wherein the old stellar population dominates as the bulk of emission from the galaxy originates from the disks. A similar observation where disks of barred galaxies were redder than the unbarred galaxies is also reported by \citet{2018Kruk}. The work suggests that the secular evolution of galaxies driven by the bar generally leads to such a scenario.

The bluer bar in the central region is associated with a radially asymmetric star-forming ring with an average (g$-$r) color $=$ 0.45 mag (Figure~\ref{fig:gr_cm}). 
Very few similar star-forming rings in S0 galaxies have been reported recently \citep{2019Proshina,2020Silchenko}. What is more intriguing is the presence of two bluer optical clumps on the ring. It is not clear how these S0 galaxies get their cold gas to drive the star formation, or could it be residual star formation activity or the UV-upturn seen in some Ellipticals \citep{OConnell1999,2007Kaviraj}.

The UV morphology of the central region of the galaxy is insightful. Figure~\ref{fig:clump} displays strong FUV and NUV emissions (observed by UVIT) from the galaxy. The FUV emission in a galaxy arises from the photosphere of young O- and B-type stars, and traces star formation up to $\sim$100 Myr \citep{kennicutt2012}. Alternatively, UV emission could arise from evolved stellar populations such as main-sequence turn-off or extreme horizontal branch stars \citep{OConnell1999}. However, the broadband UV and UV-optical colors of the galaxy are bluer, e.g., FUV$-$NUV, FUV$-$r, NUV$-$r are $\approx$ 0.66 mag, 4.15 mag, and 3.35 mag, respectively \citep{2021Pandey}, which do not seem to satisfy the UV upturn criteria \citep{2007Kaviraj,Yi_2011}. The broadband SED (Figure~\ref{fig:sed}) does not illustrate any signature of the classic UV upturn case as shown by the sample presented in \citet{Yi_2011}. 

\subsection{Star formation in clumps}
\begin{figure*}
    \centering
    \includegraphics[width=0.4\linewidth]{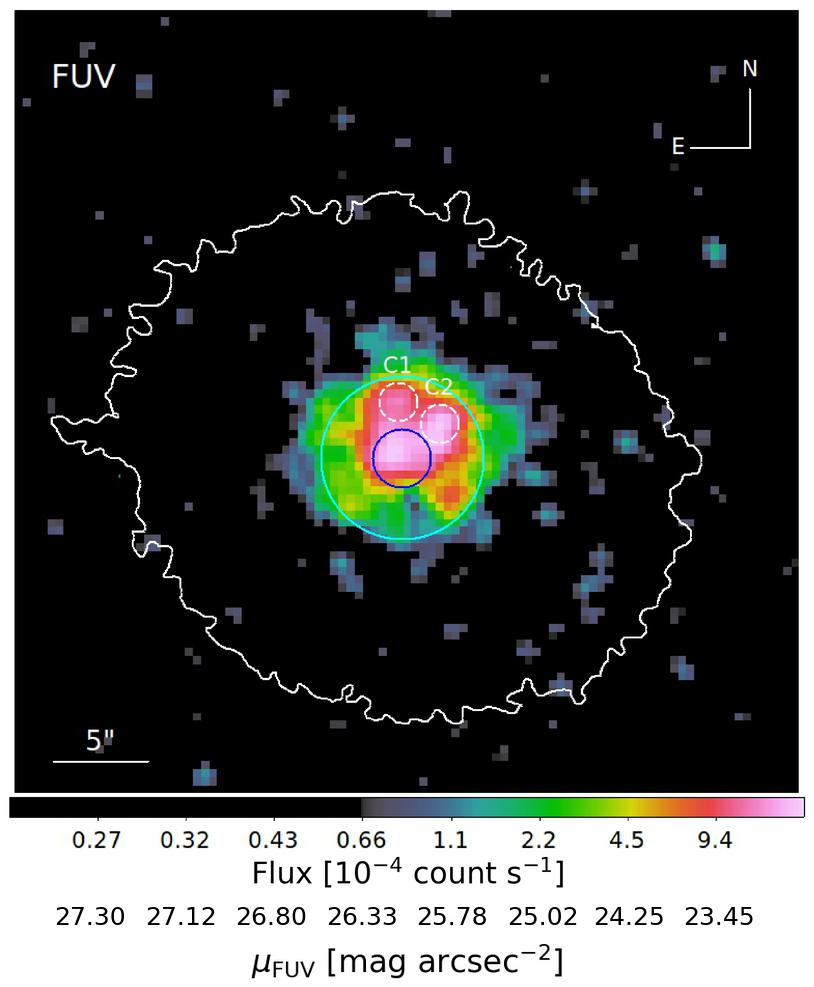}
    \includegraphics[width=0.4\linewidth]{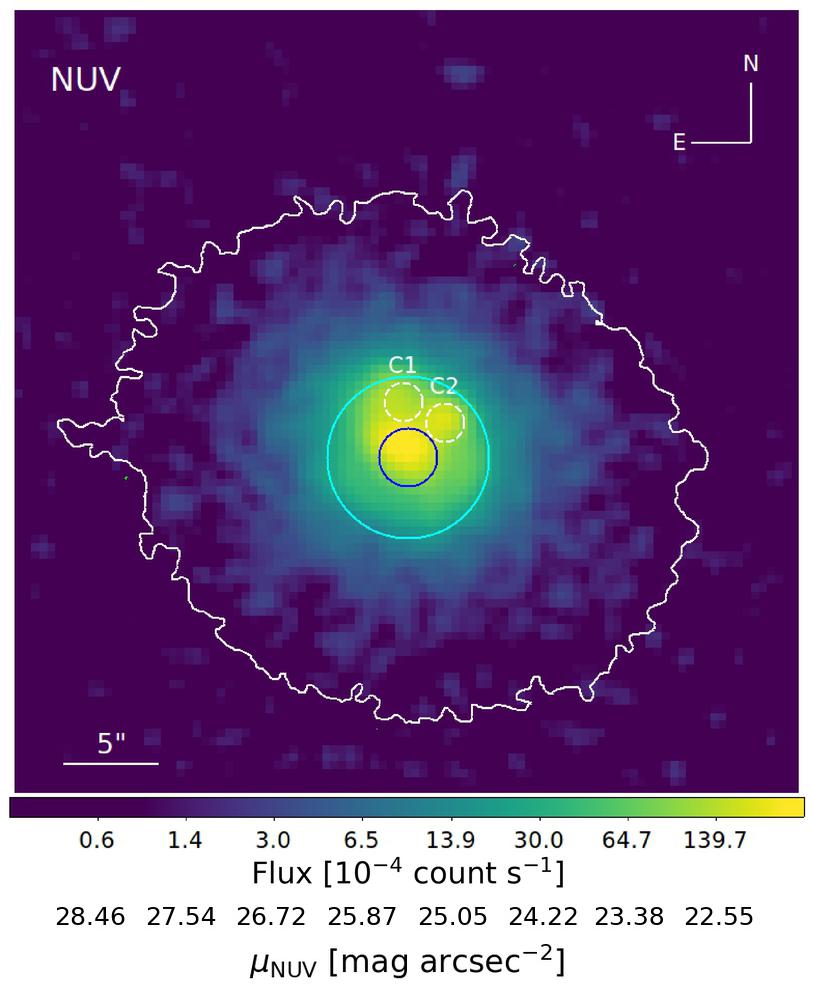}
    \caption{UVIT FUV and NUV images of the galaxy in the left and right panel, respectively. The colorbar represents the flux and surface brightness distribution on the image. The 3$\sigma$ limiting surface brightness values for FUV and NUV observation are 27.37 mag arcsec$^{-2}$ and 27.92 mag arcsec$^{-2}$, respectively. White contour represents the boundary of the galaxy seen in optical waveband, whereas blue and cyan colored circles show the central region of the galaxy (r = 1.5\arcs)
    and the outer radius of the ring (r = 4.2\arcs). The star-forming clumps C1 and C2 are enclosed in white open circle of radius 1\arcs\ each.}
    \label{fig:clump}
\end{figure*}

As seen in Figure~\ref{fig:clump}, the surface brightness of the center most region of I~Zw~81 (equivalent to aperture, r $=$ 1.5\arcs; enclosing the bar) in FUV and NUV observation are $\sim$ 23 mag arcsec$^{-2}$ and $\sim$22 mag arcsec$^{-2}$, respectively. The FUV surface brightness level at the end of the ring (r $\simeq$ $ 4.25$~kpc) is about 25~mag arcsec$^{-2}$. Beyond the extent of the outer stellar ring, the FUV emission declines sharply. Apart from the central bright UV blob, the ring at the end of the bar hosts two star-forming UV clumps (marked as C1 and C2 in Figure~\ref{fig:clump}). These clumps were identifiable in UV and optical color map of the galaxy. In a H$\alpha$ imaging survey, \citet{1993Weistrop} pointed out two small emission regions in I~Zw~81, in a direction coinciding with the positions of the UV clumps. In order to measure the FUV SFR of the clumps, we first run SExtractor on the $CFHT$ g-band image to find their central coordinates. Thereafter, we measure FUV flux around the central coordinates within a circular aperture of radius 1\arcs\ and subtract the background emission from the galaxy to get the net FUV within each clump. We calculate the FUV SFR of the clumps using the following equation \citep{1998Kennicut}:

\begin{equation}
   \rm SFR = 1.4 \times L_{FUV} (erg s^{-1} Hz^{-1}),
   \label{eq:fuvsfr}
\end{equation}

The resultant internal extinction corrected FUV SFRs for clumps C1 and C2 are 0.13 \msun yr$^{-1}$ and 0.23 \msun yr$^{-1}$, respectively. The detection of these clumps in H$\alpha$ imaging and the estimated values of FUV SFRs are a testimonial of vigorous star formation activity in clumps and not the case of UV upturn. The evidences shown above using the UV emission and optical colors are sufficient to claim that the entire central region (4$-$5 kpc) enclosing the bar is star-forming. 
\par
\begin{figure}
    \centering
    \includegraphics[width=0.95\linewidth]{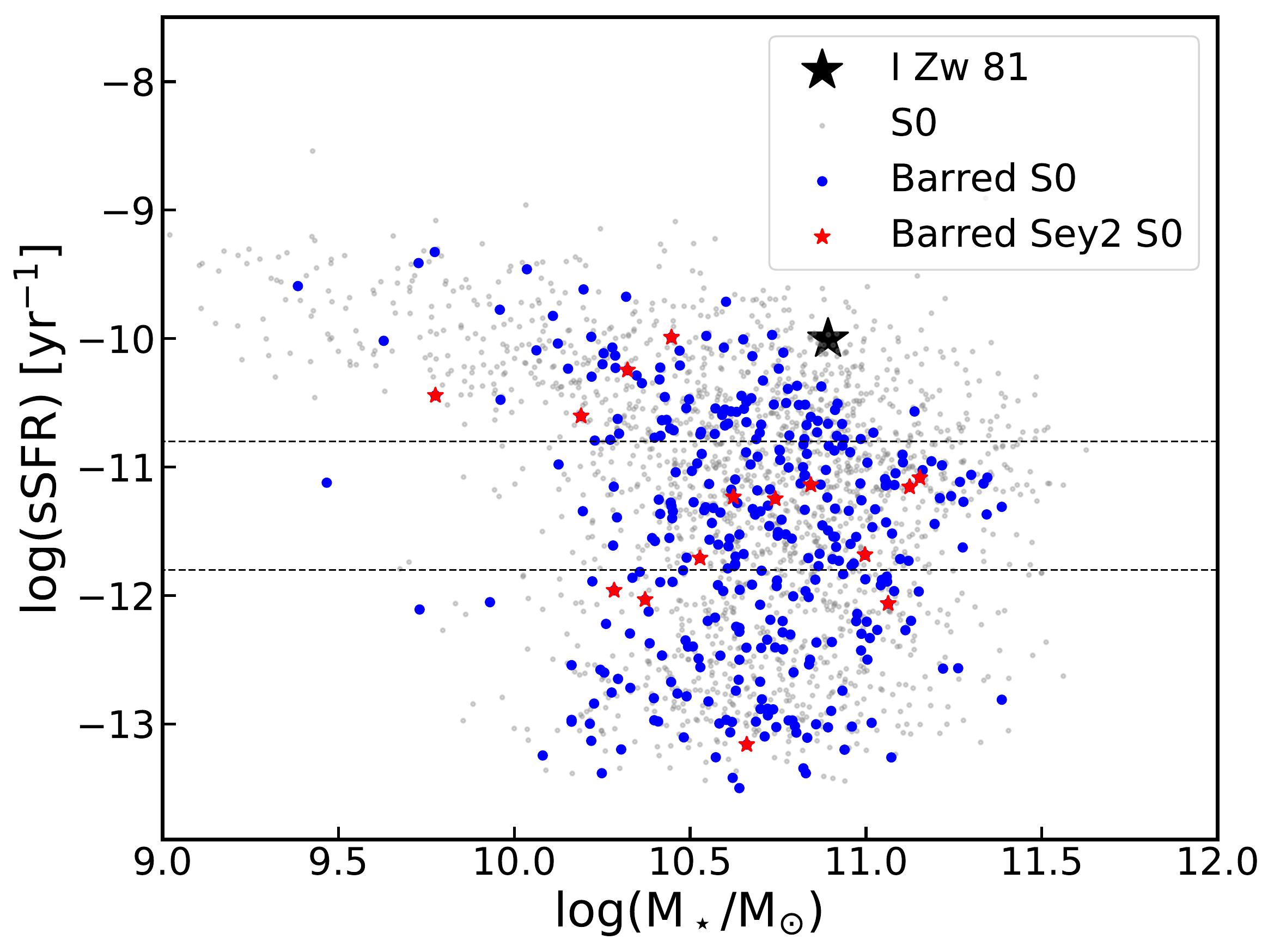}
    \caption{M$_\star$ versus sSFR distribution for a sample of S0, barred S0 and barred Sy2 S0 galaxies. The S0 sample is taken from \citet{2010Nair}.}
    \label{fig:S0s}
\end{figure}

  
To investigate whether the star-forming barred S0 galaxies are common in our local environment (z $<$ 0.1), we selected a sample of S0 (S0/a and S0) galaxies from \citet[][]{2010Nair}. The catalog sub-categorises galaxies based on their nuclear activity and marks galaxies with bars. We created a subset of S0 galaxies showing bar-like features (weak, intermediate, strong, nuclear, lens). The values of M$_\star$ and SED SFRs for the sample were taken from \citet{2016Salim} catalog. Figure~\ref{fig:S0s} shows M$_\star$ versus sSFR distribution for all (barred /barred seyfert 2) S0 galaxies alongside I~Zw~81. We notice that the frequency of star-forming barred S0 galaxies shows a gradual decrease with increasing stellar mass, and all star-forming AGN-active barred S0 galaxies are less massive than I~Zw~81.

\section{Discussion}
\label{sec:dis}
We used multi-wavelength imaging data to perform 2D structural decomposition and modeled the SED of the galaxy to understand the present evolutionary state of I~Zw~81. The disk-dominated lenticular galaxy turns out to be a star-forming with stellar mass log(M$_\star$/\msun) = 10.9. The photometric properties shown by I~Zw~81 are intriguing, e.g., shallow positive color gradient, star-forming ring, and central star formation in an old-stellar population-dominated massive galaxy. Several external and internal conditions could lead to the observed state of the galaxy, e.g., minor-mergers drive about half of the star formation activity in the local Universe \citep{2014Kaviraj} which may account for the star formation in I~Zw~81. Moreover, two internal features identified in the galaxy, i.e., AGN (with marginal evidence) and bar, are also known to affect the SFR in massive central disk galaxies similar to I~Zw~81 \citep{2021Zhang}. We discuss the impact of these processes on the galaxy.

%As a result, we primarily focus on finding if the observed state of the galaxy is a result of in-situ processes.  

\subsection{Possibility of a minor-merger interaction?}
One of the mechanisms responsible for observing a blue bar in S0 galaxies could be tidal torque and minor-mergers \citep{2020Barway}. These mechanisms were envisaged to explain the sudden bar rejuvenation seen in several S0 galaxies present in intermediate-density environment. %However, the bar-rejuvenation process as demonstrated by \citet{2020Barway} mainly occur in an  while I~Zw~81 is a void galaxy.

\begin{figure}
%    \centering
    \includegraphics[width=1.0\linewidth]{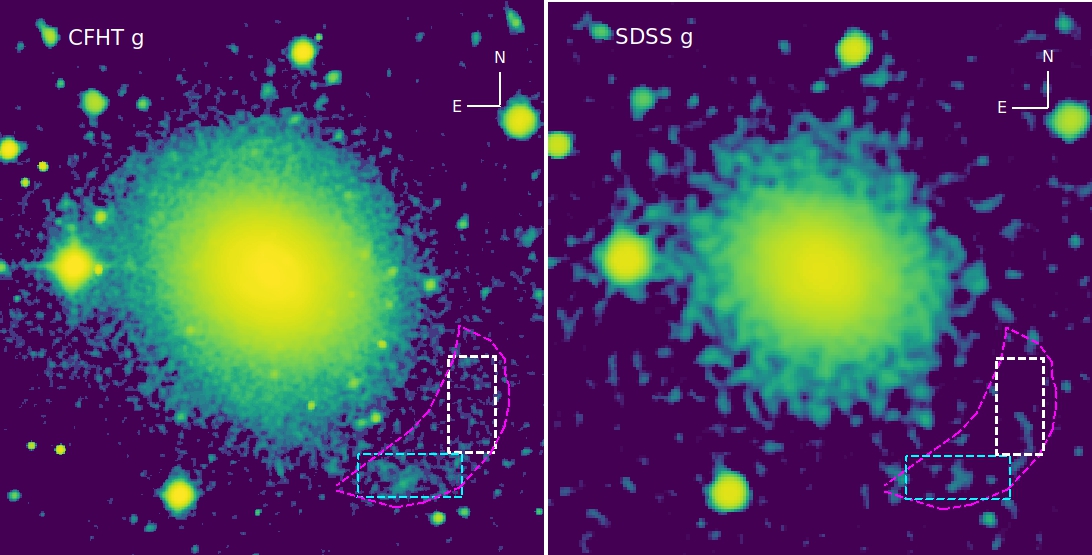}
    \caption{A comparison between $CFHT$ and $SDSS$ g-band images of I~Zw~81. The magenta-colored crescent segment, vertical and horizontal boxes  highlight the tidal tail-like feature identified in $CFHT$ observation. Same regions are overlayed on $SDSS$ g-band image.  }
    \label{fig:minor}
\end{figure}

{Based on the NED, WISEA J140811.38+485344.2, present at a distance of 0.12 Mpc, is the nearest neighbor to the galaxy. The stellar mass of WISEA J140811.38+485344.2 is $\sim$1/11 times of I Zw 81 \citep{2021Pandey}. Considering this as an ongoing interaction between the galaxies, it is viable that it could excite a bar in the host. However, the weak interaction seems insufficient to explain the observed star formation.} Thereby, we inspect $CFHT$ observations and GALFIT residues of the galaxy (Figure~\ref{fig:galfit_output}) to look for any sign of minor-merger remnants in its morphology. The completeness limit of $CFHT$ g-band observation $=$ 23.7 mag which is roughly 1.5 mag deeper than standard $SDSS$ images  \citep[magnitude limit = 22.2 mag in g-band;][]{2000York}. In fact, the depth of our $CFHT$ observation (Section~\ref{sec:skymask}) is comparable to deep $SDSS$ Stripe 82 images used by \citet{2014Kaviraj} to identify the minor mergers in local galaxies. We find a tidal tail-like feature in the observation marked by a crescent shape segment and horizontal and vertical boxes in the left panel of Figure~\ref{fig:minor}. The surface brightness within the vertical and horizontal boxes are 26.2 mag arcsec$^{-2}$ and 25.8 mag arcsec$^{-2}$, respectively. Although $SDSS$ observation (right panel of Figure~\ref{fig:minor}) do not reveal any such features, the incidence of (previous) minor merger can not be ruled out based on the $CFHT$ data. The interactions might have elevated the ongoing star formation activity in the galaxy. However, \citet{2014Kaviraj} stated that minor mergers do not significantly enhance the SFR of ETGs compared to late-type galaxies due to their low internal gas content. We need further investigation for a strong inference.

\subsubsection{Effect of interaction on the bar}
{The size of bars strongly correlates with the stellar mass of galaxies which grows in size and strength with time \citep{2019Erwin}. As a result, a massive ETG like I Zw 81 is expected to host a large bar. Instead, the detected bar is surprisingly small, with a substantial Bar/T ratio. A weak tidal interaction with the neighboring galaxy may excite the bar \citep{2019Peschken} but does not justify the observed gas influx in the bar. We argue that the small blue bar is an outcome of minor merger interactions with satellite galaxies. The young bar could be tidally induced due to recent gas accretion. Interestingly, the small bar with high SFR fits in the result presented by \citet{2020Fraser} where the scaled bar length of the galaxy correlates with its offset from the star-forming main sequence.} 

%The $CFHT$ images do not reveal any signs of morphological disturbances in the galaxy. Therefore, the star formation in I Zw 81 is less likely to be attributed to minor merger interactions. 

%Also, above a certain threshold stellar mass (log(M$_\star$) $\sim$ 10.5 \msun), evolution seems to be mainly driven by stellar mass than the environment \citep{Contini_2020}. 

\subsection{Role of AGN in central star formation}
AGNs tend to affect the evolution of galaxies in multiple ways. They spew cold gas out of the interstellar medium of the galaxy and prevent the gas from cooling down, thus leading to star formation suppression \citep[i.e., negative feedback;][]{2012Fabian}. On the other hand, AGN activity could also compress the cold and dense gas in the galaxy, which may enhance the ongoing SFR \citep[i.e., positive feedback;][]{Silk_2013}. 

Mid-IR colors are suitable for understanding the properties of active nuclei as the UV-optical radiation from the accretion disk of the AGN is absorbed by the dusty torus and re-radiated in mid-IR wavebands %\citep{2014Bagchi}
. The mid-IR colors for I~Zw~81 are $W1 - W2 = 0.3$ mag (Vega) and $W2 - W3 = 2.9$ mag (Vega). The colors infer a little re-radiated mid-IR flux from the accretion disk of the galaxy \citep{2012Smid-ir}. The result agrees with our SED analysis, where the AGN is recognized to have low luminosity. \citet{2010Shao} have reported that the star formation in the hosts of low luminosity AGNs is similar to non-AGN massive galaxies. Such hosts have an old stellar population similar to a normal ETGs \citep[D$_n$ $>$ 1.7;][]{2003Kauffman}, but I~Zw~81 possesses a younger stellar population and star formation extends further up to the outer boundary of the ring (4$-$5 kpc). Evidently, the low-luminosity AGN is insufficient in quenching the ongoing central star formation. %We require more resolved observations to probe the possibility of a positive AGN feedback in I~Zw~81.

%At some instances, the circumnuclear region (sub-kpc) of Seyfert 2 galaxies is observed as star-forming \citep[][]{2008Melendez}
\subsection{Role of bars in central star formation}
Bars are known to regulate the flow of gas to the galactic center, and enhance star-formation activity, which in turn leads to the growth of central mass concentration in galaxies \citep{Wangetal2012}. However, there is no clear picture of the morphology of star formation sites in the bar region, as many bars are devoid of star formation due to the sweeping of gas material by the bar itself \citep{2018Khoperskov}. %In some cases, there is intense star-formation activity along the inner Lindblad region (ILR) or star-formation is concentrated in the very central region \citep{Benedictetal1996, maozetal2001}.
{ Some studies found that the inner rings of barred galaxies hinder the passage of cold gas towards the galactic center by redirecting the gas in the resonant rings \citep{2019Neumann,2020Fraser}}. Massive gas-poor lenticular galaxies show FUV emission only in the central region of the bar, which is due to bar-quenching of the disks \citep{2020Diaz}. However, the FUV emission in I~Zw~81 is spread throughout the bar despite having a stellar ring (see Figure~\ref{fig:clump}). It is highlighted that only gas-rich late-type galaxies exhibit star formation along the bar \citep{2020Diaz} as shown by I~Zw~81.   %\citet{2020Fraser} have shown that  The small bar length with high SFR in I Zw 81 fits well within the picture.}

Since the bar is a strong non-axisymmetric pattern in a galaxy, it can produce enough torque to efficiently transport cold gas from the outskirts to the inner region \citep{ KormendyKennicutt2004}, excite strong star-formation activity \citep{Fanalietal2015} as well as triggers nuclear activity \citep{2021arXiv210805361S}. However, this picture of gas-inflow becomes deceptive or rather ill-understood in the case of I~Zw~81. The entire disk region in the galaxy is devoid of any spiral structure that facilitates the gas flow from the outskirts to the inner region, from where the bar torque would be more active \citep{SahaJog2014}.

I~Zw~81 is one of the few star-forming barred S0 galaxies with an active nucleus in the sample shown in Figure~\ref{fig:S0s}. The result highlights the peculiarity of the galaxy. The galaxy seems to be sufficiently gas-rich compared to other massive barred lenticular galaxies shown in Figure~\ref{fig:S0s} to support the ongoing star formation. Apart from the recent gas inflow due to minor merger, the sparse environment of the void may aid the present state of I Zw 81. Void galaxies tend to conserve their gas supplies for an extended time period due to the lack of environmental quenching processes. The combined effect perhaps lead to sSFR enhancement in the galaxy. The observed central star-formation is a result of funnelling of gas to the galactic center by the bar. A detailed study of the gas kinematics is essential to precisely understand the state of I~Zw~81.

%{\bf The presence of star formation in the ring and in the whole bar region hints towards some recent accretion of gas.} 

\section{Summary and conclusion}

We study the structural and star formation properties of a massive lenticular void galaxy - I~Zw~81. Our assimilated results are peculiar for a massive barred S0 galaxy in a void. We briefly summarize our results as follows.
\begin{itemize}
\item The galaxy has a complex morphological structure. It comprises a {NPS}, a bar, a distorted ring, a HSB inner disk, and a LSB outer disk. 
\item  Our SED analysis shows that the sSFR of the galaxy lies half a dex above the star-forming main sequence. This ongoing star formation is only limited to the central region of the galaxy.
\item The bar and inner ring displays bluer optical colors compared to the disks.
\item We detect a discernible FUV emission from the central region (4$-$5 kpc) of the galaxy. The FUV emission in I~Zw~81 is comparatively more widespread than typically seen in gas-poor S0 galaxies. The incidence of FUV emission along the bar requires the host galaxy to be gas-rich.
\item We find signature of tidal interaction in the galaxy using $CFHT$ observation. We argue that the observed SFR is due to minor merger interactions and lack of {\it galaxy nurture} in a void. {The young bar driving the gas towards the galactic center could be tidally induced due to the interaction.}
\item The low-luminosity AGN is shown to be insufficient in quenching the central star-formation.
\end{itemize}

\begin{acknowledgments}
{We thank the referee for providing constructive suggestions/comments.} This  publication  uses  the  data  from  the AstroSat mission of the Indian Space Research Organisation (ISRO) archived at the Indian Space Science Data Centre (ISSDC). ACP acknowledges the financial support of ISRO under Astrosat archival data utilization program (No. DS$\_$2B-13013(2/1/2022-Sec.2)). DP thanks Suraj Dhiwar for valuable suggestions. DP and ACP would like
to acknowledge Inter University Centre for Astronomy and Astrophysics (IUCAA), Pune, India for providing facilities to carry out this work.
\end{acknowledgments}


\vspace{5mm}
%\facilities{$CFHT$, $SDSS$, ASTROSAT/UVIT}


\software{astropy \citep{2018Astropy},  
          CIGALE \citep{2019Boquien}, 
          Source Extractor \citep{1996Bertin},
    GALFIT\citep{2010PengGF}
          }

\bibliography{sample631}{}

\begin{thebibliography}{}
\expandafter\ifx\csname natexlab\endcsname\relax\def\natexlab#1{#1}\fi
\providecommand{\url}[1]{\href{#1}{#1}}
\providecommand{\dodoi}[1]{doi:~\href{http://doi.org/#1}{\nolinkurl{#1}}}
\providecommand{\doeprint}[1]{\href{http://ascl.net/#1}{\nolinkurl{http://ascl.net/#1}}}
\providecommand{\doarXiv}[1]{\href{https://arxiv.org/abs/#1}{\nolinkurl{https://arxiv.org/abs/#1}}}

\bibitem[{{Alam} {et~al.}(2015){Alam}, {Albareti}, {Allende Prieto}, {Anders},
  {Anderson}, {Anderton}, {Andrews}, {Armengaud}, {Aubourg}, {Bailey}, {Basu},
  {Bautista}, {Beaton}, {Beers}, {Bender}, {Berlind}, {Beutler}, {Bhardwaj},
  {Bird}, {Bizyaev}, {Blake}, {Blanton}, {Blomqvist}, {Bochanski}, {Bolton},
  {Bovy}, {Shelden Bradley}, {Brandt}, {Brauer}, {Brinkmann}, {Brown},
  {Brownstein}, {Burden}, {Burtin}, {Busca}, {Cai}, {Capozzi}, {Carnero
  Rosell}, {Carr}, {Carrera}, {Chambers}, {Chaplin}, {Chen}, {Chiappini},
  {Chojnowski}, {Chuang}, {Clerc}, {Comparat}, {Covey}, {Croft}, {Cuesta},
  {Cunha}, {da Costa}, {Da Rio}, {Davenport}, {Dawson}, {De Lee}, {Delubac},
  {Deshpande}, {Dhital}, {Dutra-Ferreira}, {Dwelly}, {Ealet}, {Ebelke},
  {Edmondson}, {Eisenstein}, {Ellsworth}, {Elsworth}, {Epstein}, {Eracleous},
  {Escoffier}, {Esposito}, {Evans}, {Fan}, {Fern{\'a}ndez-Alvar}, {Feuillet},
  {Filiz Ak}, {Finley}, {Finoguenov}, {Flaherty}, {Fleming}, {Font-Ribera},
  {Foster}, {Frinchaboy}, {Galbraith-Frew}, {Garc{\'\i}a},
  {Garc{\'\i}a-Hern{\'a}ndez}, {Garc{\'\i}a P{\'e}rez}, {Gaulme}, {Ge},
  {G{\'e}nova-Santos}, {Georgakakis}, {Ghezzi}, {Gillespie}, {Girardi},
  {Goddard}, {Gontcho}, {Gonz{\'a}lez Hern{\'a}ndez}, {Grebel}, {Green},
  {Grieb}, {Grieves}, {Gunn}, {Guo}, {Harding}, {Hasselquist}, {Hawley},
  {Hayden}, {Hearty}, {Hekker}, {Ho}, {Hogg}, {Holley-Bockelmann}, {Holtzman},
  {Honscheid}, {Huber}, {Huehnerhoff}, {Ivans}, {Jiang}, {Johnson},
  {Kinemuchi}, {Kirkby}, {Kitaura}, {Klaene}, {Knapp}, {Kneib}, {Koenig},
  {Lam}, {Lan}, {Lang}, {Laurent}, {Le Goff}, {Leauthaud}, {Lee}, {Lee},
  {Licquia}, {Liu}, {Long}, {L{\'o}pez-Corredoira}, {Lorenzo-Oliveira},
  {Lucatello}, {Lundgren}, {Lupton}, {Mack}, {Mahadevan}, {Maia}, {Majewski},
  {Malanushenko}, {Malanushenko}, {Manchado}, {Manera}, {Mao}, {Maraston},
  {Marchwinski}, {Margala}, {Martell}, {Martig}, {Masters}, {Mathur},
  {McBride}, {McGehee}, {McGreer}, {McMahon}, {M{\'e}nard}, {Menzel},
  {Merloni}, {M{\'e}sz{\'a}ros}, {Miller}, {Miralda-Escud{\'e}}, {Miyatake},
  {Montero-Dorta}, {More}, {Morganson}, {Morice-Atkinson}, {Morrison},
  {Mosser}, {Muna}, {Myers}, {Nandra}, {Newman}, {Neyrinck}, {Nguyen},
  {Nichol}, {Nidever}, {Noterdaeme}, {Nuza}, {O'Connell}, {O'Connell},
  {O'Connell}, {Ogando}, {Olmstead}, {Oravetz}, {Oravetz}, {Osumi}, {Owen},
  {Padgett}, {Padmanabhan}, {Paegert}, {Palanque-Delabrouille}, {Pan},
  {Parejko}, {P{\^a}ris}, {Park}, {Pattarakijwanich}, {Pellejero-Ibanez},
  {Pepper}, {Percival}, {P{\'e}rez-Fournon}, {P{\'e}rez-R{\`a}fols},
  {Petitjean}, {Pieri}, {Pinsonneault}, {Porto de Mello}, {Prada}, {Prakash},
  {Price-Whelan}, {Protopapas}, {Raddick}, {Rahman}, {Reid}, {Rich}, {Rix},
  {Robin}, {Rockosi}, {Rodrigues}, {Rodr{\'\i}guez-Torres}, {Roe}, {Ross},
  {Ross}, {Rossi}, {Ruan}, {Rubi{\~n}o-Mart{\'\i}n}, {Rykoff},
  {Salazar-Albornoz}, {Salvato}, {Samushia}, {S{\'a}nchez}, {Santiago},
  {Sayres}, {Schiavon}, {Schlegel}, {Schmidt}, {Schneider}, {Schultheis},
  {Schwope}, {Sc{\'o}ccola}, {Scott}, {Sellgren}, {Seo}, {Serenelli}, {Shane},
  {Shen}, {Shetrone}, {Shu}, {Silva Aguirre}, {Sivarani}, {Skrutskie},
  {Slosar}, {Smith}, {Sobreira}, {Souto}, {Stassun}, {Steinmetz}, {Stello},
  {Strauss}, {Streblyanska}, {Suzuki}, {Swanson}, {Tan}, {Tayar}, {Terrien},
  {Thakar}, {Thomas}, {Thomas}, {Thompson}, {Tinker}, {Tojeiro}, {Troup},
  {Vargas-Maga{\~n}a}, {Vazquez}, {Verde}, {Viel}, {Vogt}, {Wake}, {Wang},
  {Weaver}, {Weinberg}, {Weiner}, {White}, {Wilson}, {Wisniewski},
  {Wood-Vasey}, {Ye`che}, {York}, {Zakamska}, {Zamora}, {Zasowski}, {Zehavi},
  {Zhao}, {Zheng}, {Zhou}, {Zhou}, {Zou}, \& {Zhu}}]{2015Adr12}
{Alam}, S., {Albareti}, F.~D., {Allende Prieto}, C., {et~al.} 2015, \apjs, 219,
  12

\bibitem[{{Alpaslan} {et~al.}(2015){Alpaslan}, {Driver}, {Robotham},
  {Obreschkow}, {Andrae}, {Cluver}, {Kelvin}, {Lange}, {Owers}, {Taylor},
  {Andrews}, {Bamford}, {Bland-Hawthorn}, {Brough}, {Brown}, {Colless},
  {Davies}, {Eardley}, {Grootes}, {Hopkins}, {Kennedy}, {Liske},
  {Lara-L{\'o}pez}, {L{\'o}pez-S{\'a}nchez}, {Loveday}, {Madore}, {Mahajan},
  {Meyer}, {Moffett}, {Norberg}, {Penny}, {Pimbblet}, {Popescu}, {Seibert}, \&
  {Tuffs}}]{2015Alpaslan}
{Alpaslan}, M., {Driver}, S., {Robotham}, A. S.~G., {et~al.} 2015, \mnras, 451,
  3249

\bibitem[{{Astropy Collaboration} {et~al.}(2018){Astropy Collaboration},
  {Price-Whelan}, {Sip{\H{o}}cz}, {G{\"u}nther}, {Lim}, {Crawford}, {Conseil},
  {Shupe}, {Craig}, {Dencheva}, {Ginsburg}, {VanderPlas}, {Bradley},
  {P{\'e}rez-Su{\'a}rez}, {de Val-Borro}, {Aldcroft}, {Cruz}, {Robitaille},
  {Tollerud}, {Ardelean}, {Babej}, {Bach}, {Bachetti}, {Bakanov}, {Bamford},
  {Barentsen}, {Barmby}, {Baumbach}, {Berry}, {Biscani}, {Boquien}, {Bostroem},
  {Bouma}, {Brammer}, {Bray}, {Breytenbach}, {Buddelmeijer}, {Burke},
  {Calderone}, {Cano Rodr{\'\i}guez}, {Cara}, {Cardoso}, {Cheedella}, {Copin},
  {Corrales}, {Crichton}, {D'Avella}, {Deil}, {Depagne}, {Dietrich}, {Donath},
  {Droettboom}, {Earl}, {Erben}, {Fabbro}, {Ferreira}, {Finethy}, {Fox},
  {Garrison}, {Gibbons}, {Goldstein}, {Gommers}, {Greco}, {Greenfield},
  {Groener}, {Grollier}, {Hagen}, {Hirst}, {Homeier}, {Horton}, {Hosseinzadeh},
  {Hu}, {Hunkeler}, {Ivezi{\'c}}, {Jain}, {Jenness}, {Kanarek}, {Kendrew},
  {Kern}, {Kerzendorf}, {Khvalko}, {King}, {Kirkby}, {Kulkarni}, {Kumar},
  {Lee}, {Lenz}, {Littlefair}, {Ma}, {Macleod}, {Mastropietro}, {McCully},
  {Montagnac}, {Morris}, {Mueller}, {Mumford}, {Muna}, {Murphy}, {Nelson},
  {Nguyen}, {Ninan}, {N{\"o}the}, {Ogaz}, {Oh}, {Parejko}, {Parley}, {Pascual},
  {Patil}, {Patil}, {Plunkett}, {Prochaska}, {Rastogi}, {Reddy Janga},
  {Sabater}, {Sakurikar}, {Seifert}, {Sherbert}, {Sherwood-Taylor}, {Shih},
  {Sick}, {Silbiger}, {Singanamalla}, {Singer}, {Sladen}, {Sooley},
  {Sornarajah}, {Streicher}, {Teuben}, {Thomas}, {Tremblay}, {Turner},
  {Terr{\'o}n}, {van Kerkwijk}, {de la Vega}, {Watkins}, {Weaver}, {Whitmore},
  {Woillez}, {Zabalza}, \& {Astropy Contributors}}]{2018Astropy}
{Astropy Collaboration}, {Price-Whelan}, A.~M., {Sip{\H{o}}cz}, B.~M., {et~al.}
  2018, \aj, 156, 123

\bibitem[{{Balogh} {et~al.}(2004){Balogh}, {Baldry}, {Nichol}, {Miller},
  {Bower}, \& {Glazebrook}}]{2004Balogh}
{Balogh}, M.~L., {Baldry}, I.~K., {Nichol}, R., {et~al.} 2004, \apjl, 615, L101

\bibitem[{{Balogh} {et~al.}(1999){Balogh}, {Morris}, {Yee}, {Carlberg}, \&
  {Ellingson}}]{1999Balogh}
{Balogh}, M.~L., {Morris}, S.~L., {Yee}, H.~K.~C., {Carlberg}, R.~G., \&
  {Ellingson}, E. 1999, \apj, 527, 54

\bibitem[{{Barway} \& {Saha}(2020)}]{2020Barway}
{Barway}, S., \& {Saha}, K. 2020, \mnras, 495, 4548

\bibitem[{{Barway} {et~al.}(2013){Barway}, {Wadadekar}, {Vaghmare}, \&
  {Kembhavi}}]{2013Barway}
{Barway}, S., {Wadadekar}, Y., {Vaghmare}, K., \& {Kembhavi}, A.~K. 2013,
  \mnras, 432, 430

\bibitem[{{Bertin} \& {Arnouts}(1996)}]{1996Bertin}
{Bertin}, E., \& {Arnouts}, S. 1996, \aaps, 117, 393

\bibitem[{{Beygu} {et~al.}(2017){Beygu}, {Peletier}, {van der Hulst},
  {Jarrett}, {Kreckel}, {van de Weygaert}, {van Gorkom}, \&
  {Aragon-Calvo}}]{2017Beygu}
{Beygu}, B., {Peletier}, R.~F., {van der Hulst}, J.~M., {et~al.} 2017, \mnras,
  464, 666

\bibitem[{Bluck {et~al.}(2020)Bluck, Maiolino, Piotrowska, Trussler, Ellison,
  Sánchez, Thorp, Teimoorinia, Moreno, \& Conselice}]{Bluck2020}
Bluck, A. F.~L., Maiolino, R., Piotrowska, J.~M., {et~al.} 2020, \mnras, 499,
  230

\bibitem[{{Boquien} {et~al.}(2019){Boquien}, {Burgarella}, {Roehlly}, {Buat},
  {Ciesla}, {Corre}, {Inoue}, \& {Salas}}]{2019Boquien}
{Boquien}, M., {Burgarella}, D., {Roehlly}, Y., {et~al.} 2019, \aap, 622, A103

\bibitem[{{Boulade} {et~al.}(1998){Boulade}, {Vigroux}, {Charlot}, {Borgeaud},
  {Carton}, {de Kat}, {Rousse}, {Mellier}, {Gigan}, {Crampton}, \&
  {Morbey}}]{CFHT}
{Boulade}, O., {Vigroux}, L.~G., {Charlot}, X., {et~al.} 1998, in Society of
  Photo-Optical Instrumentation Engineers (SPIE) Conference Series, Vol. 3355,
  Optical Astronomical Instrumentation, ed. S.~{D'Odorico}, 614--625

\bibitem[{Bradley {et~al.}(2020)Bradley, Sip{\H o}cz, Robitaille, Tollerud,
  Vin{\'{\i}}cius, Deil, Barbary, Wilson, Busko, G{\"u}nther, Cara, Conseil,
  Bostroem, Droettboom, Bray, Bratholm, Lim, Barentsen, Craig, Pascual, Perren,
  Greco, Donath, de~Val-Borro, Kerzendorf, Bach, Weaver, D'Eugenio, Souchereau,
  \& Ferreira}]{bradley_2020}
Bradley, L., Sip{\H o}cz, B., Robitaille, T., {et~al.} 2020, astropy/photutils:
  1.0.0, 1.0.0,  Zenodo

\bibitem[{{Bruzual} \& {Charlot}(2003)}]{2003Bruzual}
{Bruzual}, G., \& {Charlot}, S. 2003, \mnras, 344, 1000

\bibitem[{{Buta} {et~al.}(2010){Buta}, {Laurikainen}, {Salo}, \&
  {Knapen}}]{2010Buta}
{Buta}, R., {Laurikainen}, E., {Salo}, H., \& {Knapen}, J.~H. 2010, \apj, 721,
  259

\bibitem[{{Calzetti} {et~al.}(2000){Calzetti}, {Armus}, {Bohlin}, {Kinney},
  {Koornneef}, \& {Storchi-Bergmann}}]{2000Calzetti}
{Calzetti}, D., {Armus}, L., {Bohlin}, R.~C., {et~al.} 2000, \apj, 533, 682

\bibitem[{{Chilingarian} \& {Zolotukhin}(2012)}]{2012Chiligan}
{Chilingarian}, I.~V., \& {Zolotukhin}, I.~Y. 2012, \mnras, 419, 1727

\bibitem[{{Ciesla} {et~al.}(2018){Ciesla}, {Elbaz}, {Schreiber}, {Daddi}, \&
  {Wang}}]{2018Ciesla}
{Ciesla}, L., {Elbaz}, D., {Schreiber}, C., {Daddi}, E., \& {Wang}, T. 2018,
  \aap, 615, A61

\bibitem[{{Constantin} {et~al.}(2008){Constantin}, {Hoyle}, \&
  {Vogeley}}]{2008Constantin}
{Constantin}, A., {Hoyle}, F., \& {Vogeley}, M.~S. 2008, \apj, 673, 715

\bibitem[{{Cooper} {et~al.}(2007){Cooper}, {Newman}, {Coil}, {Croton}, {Gerke},
  {Yan}, {Davis}, {Faber}, {Guhathakurta}, {Koo}, {Weiner}, \&
  {Willmer}}]{2007Cooper}
{Cooper}, M.~C., {Newman}, J.~A., {Coil}, A.~L., {et~al.} 2007, \mnras, 376,
  1445

\bibitem[{{Croton} {et~al.}(2006){Croton}, {Springel}, {White}, {De Lucia},
  {Frenk}, {Gao}, {Jenkins}, {Kauffmann}, {Navarro}, \& {Yoshida}}]{2006Croton}
{Croton}, D.~J., {Springel}, V., {White}, S. D.~M., {et~al.} 2006, \mnras, 365,
  11

\bibitem[{{Cruzen} {et~al.}(2002){Cruzen}, {Wehr}, {Weistrop}, {Angione}, \&
  {Hoopes}}]{2002Cruzen}
{Cruzen}, S., {Wehr}, T., {Weistrop}, D., {Angione}, R.~J., \& {Hoopes}, C.
  2002, \aj, 123, 142

\bibitem[{{Cruzen} {et~al.}(1997){Cruzen}, {Weistrop}, \&
  {Hoopes}}]{1997Cruzen}
{Cruzen}, S.~T., {Weistrop}, D., \& {Hoopes}, C.~G. 1997, \aj, 113, 1983

\bibitem[{{Cutri} {et~al.}(2013){Cutri}, {Wright}, {Conrow}, {Fowler},
  {Eisenhardt}, {Grillmair}, {Kirkpatrick}, {Masci}, {McCallon}, {Wheelock},
  {Fajardo-Acosta}, {Yan}, {Benford}, {Harbut}, {Jarrett}, {Lake}, {Leisawitz},
  {Ressler}, {Stanford}, {Tsai}, {Liu}, {Helou}, {Mainzer}, {Gettings},
  {Gonzalez}, {Hoffman}, {Marsh}, {Padgett}, {Skrutskie}, {Beck}, {Papin}, \&
  {Wittman}}]{2013wise}
{Cutri}, R.~M., {Wright}, E.~L., {Conrow}, T., {et~al.} 2013, {Explanatory
  Supplement to the AllWISE Data Release Products}, Explanatory Supplement to
  the AllWISE Data Release Products

\bibitem[{{Dale} {et~al.}(2014){Dale}, {Helou}, {Magdis}, {Armus},
  {D{\'\i}az-Santos}, \& {Shi}}]{2014Dale}
{Dale}, D.~A., {Helou}, G., {Magdis}, G.~E., {et~al.} 2014, \apj, 784, 83

\bibitem[{{Dekel} \& {Birnboim}(2006)}]{2006Dekelandb}
{Dekel}, A., \& {Birnboim}, Y. 2006, \mnras, 368, 2

\bibitem[{{D{\'\i}az-Garc{\'\i}a} {et~al.}(2020){D{\'\i}az-Garc{\'\i}a},
  {Moyano}, {Comer{\'o}n}, {Knapen}, {Salo}, \& {Bouquin}}]{2020Diaz}
{D{\'\i}az-Garc{\'\i}a}, S., {Moyano}, F.~D., {Comer{\'o}n}, S., {et~al.} 2020,
  \aap, 644, A38

\bibitem[{{Dressler}(1980)}]{1980Dressler}
{Dressler}, A. 1980, \apj, 236, 351

\bibitem[{{Eliche-Moral} {et~al.}(2018){Eliche-Moral},
  {Rodr{\'\i}guez-P{\'e}rez}, {Borlaff}, {Querejeta}, \& {Tapia}}]{2018moral}
{Eliche-Moral}, M.~C., {Rodr{\'\i}guez-P{\'e}rez}, C., {Borlaff}, A.,
  {Querejeta}, M., \& {Tapia}, T. 2018, \aap, 617, A113

\bibitem[{{Erwin}(2019)}]{2019Erwin}
{Erwin}, P. 2019, \mnras, 489, 3553

\bibitem[{{Fabian}(2012)}]{2012Fabian}
{Fabian}, A.~C. 2012, \araa, 50, 455

\bibitem[{{Fanali} {et~al.}(2015){Fanali}, {Dotti}, {Fiacconi}, \&
  {Haardt}}]{Fanalietal2015}
{Fanali}, R., {Dotti}, M., {Fiacconi}, D., \& {Haardt}, F. 2015, \mnras, 454,
  3641

\bibitem[{{Fisher} \& {Drory}(2008)}]{2008Fisher}
{Fisher}, D.~B., \& {Drory}, N. 2008, \aj, 136, 773

\bibitem[{Fraser-McKelvie {et~al.}(2016)Fraser-McKelvie, Pimbblet, Penny, \&
  Brown}]{2016mckelvie}
Fraser-McKelvie, A., Pimbblet, K.~A., Penny, S.~J., \& Brown, M. J.~I. 2016,
  \mnras, 459, 754

\bibitem[{{Fraser-McKelvie} {et~al.}(2020){Fraser-McKelvie},
  {Arag{\'o}n-Salamanca}, {Merrifield}, {Masters}, {Nair}, {Emsellem},
  {Kraljic}, {Krishnarao}, {Andrews}, {Drory}, \& {Neumann}}]{2020Fraser}
{Fraser-McKelvie}, A., {Arag{\'o}n-Salamanca}, A., {Merrifield}, M., {et~al.}
  2020, \mnras, 495, 4158

\bibitem[{{Genzel} {et~al.}(2014){Genzel}, {F{\"o}rster Schreiber}, {Lang},
  {Tacchella}, {Tacconi}, {Wuyts}, {Bandara}, {Burkert}, {Buschkamp},
  {Carollo}, {Cresci}, {Davies}, {Eisenhauer}, {Hicks}, {Kurk}, {Lilly},
  {Lutz}, {Mancini}, {Naab}, {Newman}, {Peng}, {Renzini}, {Shapiro Griffin},
  {Sternberg}, {Vergani}, {Wisnioski}, {Wuyts}, \& {Zamorani}}]{2014Genzel}
{Genzel}, R., {F{\"o}rster Schreiber}, N.~M., {Lang}, P., {et~al.} 2014, \apj,
  785, 75

\bibitem[{{Grogin} \& {Geller}(1999)}]{1999Grogin}
{Grogin}, N.~A., \& {Geller}, M.~J. 1999, \aj, 118, 2561

\bibitem[{{Gunn} \& {Gott}(1972)}]{1972Gunn}
{Gunn}, J.~E., \& {Gott}, J.~Richard, I. 1972, \apj, 176, 1

\bibitem[{{Haines} {et~al.}(2007){Haines}, {Gargiulo}, {La Barbera},
  {Mercurio}, {Merluzzi}, \& {Busarello}}]{2007Haines}
{Haines}, C.~P., {Gargiulo}, A., {La Barbera}, F., {et~al.} 2007, \mnras, 381,
  7

\bibitem[{{Kannappan} {et~al.}(2009){Kannappan}, {Guie}, \&
  {Baker}}]{2009kannappan}
{Kannappan}, S.~J., {Guie}, J.~M., \& {Baker}, A.~J. 2009, \aj, 138, 579

\bibitem[{Kauffmann {et~al.}(2004)Kauffmann, White, Heckman, Ménard,
  Brinchmann, Charlot, Tremonti, \& Brinkmann}]{Kauffmann2004}
Kauffmann, G., White, S. D.~M., Heckman, T.~M., {et~al.} 2004, \mnras, 353, 713

\bibitem[{{Kauffmann} {et~al.}(2003{\natexlab{a}}){Kauffmann}, {Heckman},
  {White}, {Charlot}, {Tremonti}, {Peng}, {Seibert}, {Brinkmann}, {Nichol},
  {SubbaRao}, \& {York}}]{2003KauffmannM}
{Kauffmann}, G., {Heckman}, T.~M., {White}, S. D.~M., {et~al.}
  2003{\natexlab{a}}, \mnras, 341, 54

\bibitem[{{Kauffmann} {et~al.}(2003{\natexlab{b}}){Kauffmann}, {Heckman},
  {Tremonti}, {Brinchmann}, {Charlot}, {White}, {Ridgway}, {Brinkmann},
  {Fukugita}, {Hall}, {Ivezi{\'c}}, {Richards}, \& {Schneider}}]{2003Kauffman}
{Kauffmann}, G., {Heckman}, T.~M., {Tremonti}, C., {et~al.} 2003{\natexlab{b}},
  \mnras, 346, 1055

\bibitem[{{Kaviraj}(2014)}]{2014Kaviraj}
{Kaviraj}, S. 2014, \mnras, 440, 2944

\bibitem[{{Kaviraj} {et~al.}(2007){Kaviraj}, {Schawinski}, {Devriendt},
  {Ferreras}, {Khochfar}, {Yoon}, {Yi}, {Deharveng}, {Boselli}, {Barlow},
  {Conrow}, {Forster}, {Friedman}, {Martin}, {Morrissey}, {Neff},
  {Schiminovich}, {Seibert}, {Small}, {Wyder}, {Bianchi}, {Donas}, {Heckman},
  {Lee}, {Madore}, {Milliard}, {Rich}, \& {Szalay}}]{2007Kaviraj}
{Kaviraj}, S., {Schawinski}, K., {Devriendt}, J.~E.~G., {et~al.} 2007, \apjs,
  173, 619

\bibitem[{{Kennicutt}(1998)}]{1998Kennicut}
{Kennicutt}, Robert~C., J. 1998, \araa, 36, 189

\bibitem[{Kennicutt \& Evans(2012)}]{kennicutt2012}
Kennicutt, R.~C., \& Evans, N.~J. 2012, \araa, 50, 531

\bibitem[{{Khoperskov} {et~al.}(2018){Khoperskov}, {Haywood}, {Di Matteo},
  {Lehnert}, \& {Combes}}]{2018Khoperskov}
{Khoperskov}, S., {Haywood}, M., {Di Matteo}, P., {Lehnert}, M.~D., \&
  {Combes}, F. 2018, \aap, 609, A60

\bibitem[{{Kormendy}(2016)}]{kormendy2016}
{Kormendy}, J. 2016, in Astrophysics and Space Science Library, Vol. 418,
  Galactic Bulges, ed. E.~{Laurikainen}, R.~{Peletier}, \& D.~{Gadotti}, 431

\bibitem[{{Kormendy} \& {Kennicutt}(2004)}]{KormendyKennicutt2004}
{Kormendy}, J., \& {Kennicutt}, Robert~C., J. 2004, \araa, 42, 603

\bibitem[{{Kreckel} {et~al.}(2012){Kreckel}, {Platen}, {Arag{\'o}n-Calvo}, {van
  Gorkom}, {van de Weygaert}, {van der Hulst}, \& {Beygu}}]{2012Kreckel}
{Kreckel}, K., {Platen}, E., {Arag{\'o}n-Calvo}, M.~A., {et~al.} 2012, \aj,
  144, 16

\bibitem[{{Kruk} {et~al.}(2018){Kruk}, {Lintott}, {Bamford}, {Masters},
  {Simmons}, {H{\"a}u{\ss}ler}, {Cardamone}, {Hart}, {Kelvin}, {Schawinski},
  {Smethurst}, \& {Vika}}]{2018Kruk}
{Kruk}, S.~J., {Lintott}, C.~J., {Bamford}, S.~P., {et~al.} 2018, \mnras, 473,
  4731

\bibitem[{{Lipovetsky} {et~al.}(1988){Lipovetsky}, {Neizvestny}, \&
  {Neizvestnaya}}]{1988seyfert}
{Lipovetsky}, V.~A., {Neizvestny}, S.~I., \& {Neizvestnaya}, O.~M. 1988,
  Soobshcheniya Spetsial'noj Astrofizicheskoj Observatorii, 55, 5

\bibitem[{{Liske} {et~al.}(2015){Liske}, {Baldry}, {Driver}, {Tuffs},
  {Alpaslan}, {Andrae}, {Brough}, {Cluver}, {Grootes}, {Gunawardhana},
  {Kelvin}, {Loveday}, {Robotham}, {Taylor}, {Bamford}, {Bland-Hawthorn},
  {Brown}, {Drinkwater}, {Hopkins}, {Meyer}, {Norberg}, {Peacock}, {Agius},
  {Andrews}, {Bauer}, {Ching}, {Colless}, {Conselice}, {Croom}, {Davies}, {De
  Propris}, {Dunne}, {Eardley}, {Ellis}, {Foster}, {Frenk}, {H{\"a}u{\ss}ler},
  {Holwerda}, {Howlett}, {Ibarra}, {Jarvis}, {Jones}, {Kafle}, {Lacey},
  {Lange}, {Lara-L{\'o}pez}, {L{\'o}pez-S{\'a}nchez}, {Maddox}, {Madore},
  {McNaught-Roberts}, {Moffett}, {Nichol}, {Owers}, {Palamara}, {Penny},
  {Phillipps}, {Pimbblet}, {Popescu}, {Prescott}, {Proctor}, {Sadler},
  {Sansom}, {Seibert}, {Sharp}, {Sutherland}, {V{\'a}zquez-Mata}, {van Kampen},
  {Wilkins}, {Williams}, \& {Wright}}]{2015gama}
{Liske}, J., {Baldry}, I.~K., {Driver}, S.~P., {et~al.} 2015, \mnras, 452, 2087

\bibitem[{{Maiolino} {et~al.}(2012){Maiolino}, {Gallerani}, {Neri}, {Cicone},
  {Ferrara}, {Genzel}, {Lutz}, {Sturm}, {Tacconi}, {Walter}, {Feruglio},
  {Fiore}, \& {Piconcelli}}]{2012Maiolino}
{Maiolino}, R., {Gallerani}, S., {Neri}, R., {et~al.} 2012, \mnras, 425, L66

\bibitem[{{Martig} {et~al.}(2009){Martig}, {Bournaud}, {Teyssier}, \&
  {Dekel}}]{2009Martig}
{Martig}, M., {Bournaud}, F., {Teyssier}, R., \& {Dekel}, A. 2009, \apj, 707,
  250

\bibitem[{{Moore} {et~al.}(1996){Moore}, {Katz}, {Lake}, {Dressler}, \&
  {Oemler}}]{1996Galhar}
{Moore}, B., {Katz}, N., {Lake}, G., {Dressler}, A., \& {Oemler}, A. 1996,
  \nat, 379, 613

\bibitem[{{Moshir} {et~al.}(1990){Moshir}, {Kopan}, {Conrow}, {McCallon},
  {Hacking}, {Gregorich}, {Rohrbach}, {Melnyk}, {Rice}, {Fullmer}, {White}, \&
  {Chester}}]{1990Moshir}
{Moshir}, M., {Kopan}, G., {Conrow}, T., {et~al.} 1990, in Bulletin of the
  American Astronomical Society, Vol.~22, 1325

\bibitem[{{Nair} \& {Abraham}(2010)}]{2010Nair}
{Nair}, P.~B., \& {Abraham}, R.~G. 2010, \apjs, 186, 427

\bibitem[{{Neumann} {et~al.}(2017){Neumann}, {Wisotzki}, {Choudhury},
  {Gadotti}, {Walcher}, {Bland-Hawthorn}, {Garc{\'\i}a-Benito}, {Gonz{\'a}lez
  Delgado}, {Husemann}, {Marino}, {M{\'a}rquez}, {S{\'a}nchez}, {Ziegler}, \&
  {CALIFA Collaboration}}]{2017Neumann}
{Neumann}, J., {Wisotzki}, L., {Choudhury}, O.~S., {et~al.} 2017, \aap, 604,
  A30

\bibitem[{{Neumann} {et~al.}(2019){Neumann}, {Gadotti}, {Wisotzki}, {Husemann},
  {Busch}, {Combes}, {Croom}, {Davis}, {Gaspari}, {Krumpe}, {P{\'e}rez-Torres},
  {Scharw{\"a}chter}, {Smirnova-Pinchukova}, {Tremblay}, \&
  {Urrutia}}]{2019Neumann}
{Neumann}, J., {Gadotti}, D.~A., {Wisotzki}, L., {et~al.} 2019, \aap, 627, A26

\bibitem[{{O'Connell}(1999)}]{OConnell1999}
{O'Connell}, R.~W. 1999, \araa, 37, 603

\bibitem[{{Oke} \& {Gunn}(1983)}]{1983Oke}
{Oke}, J.~B., \& {Gunn}, J.~E. 1983, \apj, 266, 713

\bibitem[{{Osterbrock} \& {Ferland}(2006)}]{2006Osterbrock}
{Osterbrock}, D.~E., \& {Ferland}, G.~J. 2006, {Astrophysics of gaseous nebulae
  and active galactic nuclei}

\bibitem[{Pan {et~al.}(2012)Pan, Vogeley, Hoyle, Choi, \& Park}]{pan2012}
Pan, D.~C., Vogeley, M.~S., Hoyle, F., Choi, Y.-Y., \& Park, C. 2012, \mnras,
  421, 926

\bibitem[{{Pandey} {et~al.}(2021){Pandey}, {Saha}, \& {Pradhan}}]{2021Pandey}
{Pandey}, D., {Saha}, K., \& {Pradhan}, A.~C. 2021, \apj, 919, 101

\bibitem[{{Peng} {et~al.}(2002){Peng}, {Ho}, {Impey}, \& {Rix}}]{2002Peng}
{Peng}, C.~Y., {Ho}, L.~C., {Impey}, C.~D., \& {Rix}, H.-W. 2002, \aj, 124, 266

\bibitem[{{Peng} {et~al.}(2010{\natexlab{a}}){Peng}, {Ho}, {Impey}, \&
  {Rix}}]{2010PengGF}
---. 2010{\natexlab{a}}, \aj, 139, 2097

\bibitem[{{Peng} {et~al.}(2010{\natexlab{b}}){Peng}, {Lilly}, {Kova{\v{c}}},
  {Bolzonella}, {Pozzetti}, {Renzini}, {Zamorani}, {Ilbert}, {Knobel},
  {Iovino}, {Maier}, {Cucciati}, {Tasca}, {Carollo}, {Silverman}, {Kampczyk},
  {de Ravel}, {Sanders}, {Scoville}, {Contini}, {Mainieri}, {Scodeggio},
  {Kneib}, {Le F{\`e}vre}, {Bardelli}, {Bongiorno}, {Caputi}, {Coppa}, {de la
  Torre}, {Franzetti}, {Garilli}, {Lamareille}, {Le Borgne}, {Le Brun},
  {Mignoli}, {Perez Montero}, {Pello}, {Ricciardelli}, {Tanaka}, {Tresse},
  {Vergani}, {Welikala}, {Zucca}, {Oesch}, {Abbas}, {Barnes}, {Bordoloi},
  {Bottini}, {Cappi}, {Cassata}, {Cimatti}, {Fumana}, {Hasinger}, {Koekemoer},
  {Leauthaud}, {Maccagni}, {Marinoni}, {McCracken}, {Memeo}, {Meneux}, {Nair},
  {Porciani}, {Presotto}, \& {Scaramella}}]{2010Peng}
{Peng}, Y.-j., {Lilly}, S.~J., {Kova{\v{c}}}, K., {et~al.} 2010{\natexlab{b}},
  \apj, 721, 193

\bibitem[{Penny {et~al.}(2015)Penny, Brown, Pimbblet, Cluver, Croton, Owers,
  Lange, Alpaslan, Baldry, Bland-Hawthorn, Brough, Driver, Holwerda, Hopkins,
  Jarrett, Jones, Kelvin, Lara-López, Liske, López-Sánchez, Loveday, Meyer,
  Norberg, Robotham, \& Rodrigues}]{SJPenny}
Penny, S.~J., Brown, M. J.~I., Pimbblet, K.~A., {et~al.} 2015, \mnras, 453,
  3519

\bibitem[{{Peschken} \& {{\L}okas}(2019)}]{2019Peschken}
{Peschken}, N., \& {{\L}okas}, E.~L. 2019, \mnras, 483, 2721

\bibitem[{{Pettini} \& {Pagel}(2004)}]{2004Pettini}
{Pettini}, M., \& {Pagel}, B. E.~J. 2004, \mnras, 348, L59

\bibitem[{{Proshina} {et~al.}(2019){Proshina}, {Kniazev}, \&
  {Sil'chenko}}]{2019Proshina}
{Proshina}, I.~S., {Kniazev}, A.~Y., \& {Sil'chenko}, O.~K. 2019, \aj, 158, 5

\bibitem[{{Pustilnik} {et~al.}(2011){Pustilnik}, {Tepliakova}, \&
  {Kniazev}}]{2011PT}
{Pustilnik}, S.~A., {Tepliakova}, A.~L., \& {Kniazev}, A.~Y. 2011,
  Astrophysical Bulletin, 66, 255

\bibitem[{{Richstone}(1976)}]{1976Richstone}
{Richstone}, D.~O. 1976, \apj, 204, 642

\bibitem[{{Rojas} {et~al.}(2004){Rojas}, {Vogeley}, {Hoyle}, \&
  {Brinkmann}}]{2004Rojas}
{Rojas}, R.~R., {Vogeley}, M.~S., {Hoyle}, F., \& {Brinkmann}, J. 2004, \apj,
  617, 50

\bibitem[{{Rosenbaum} {et~al.}(2009){Rosenbaum}, {Krusch}, {Bomans}, \&
  {Dettmar}}]{2009Rosenbaum}
{Rosenbaum}, S.~D., {Krusch}, E., {Bomans}, D.~J., \& {Dettmar}, R.~J. 2009,
  \aap, 504, 807

\bibitem[{{Saha}(2015)}]{2015Saha}
{Saha}, K. 2015, \apjl, 806, L29

\bibitem[{{Saha} {et~al.}(2021){Saha}, {Dhiwar}, {Barway}, {Narayan}, \&
  {Tandon}}]{Sahaetal2021}
{Saha}, K., {Dhiwar}, S., {Barway}, S., {Narayan}, C., \& {Tandon}, S. 2021,
  Journal of Astrophysics and Astronomy, 42, 59

\bibitem[{{Saha} \& {Jog}(2014)}]{SahaJog2014}
{Saha}, K., \& {Jog}, C.~J. 2014, \mnras, 444, 352

\bibitem[{{Saha} {et~al.}(2012){Saha}, {Martinez-Valpuesta}, \&
  {Gerhard}}]{2012Sahaetal}
{Saha}, K., {Martinez-Valpuesta}, I., \& {Gerhard}, O. 2012, \mnras, 421, 333

\bibitem[{{Salim} {et~al.}(2016){Salim}, {Lee}, {Janowiecki}, {da Cunha},
  {Dickinson}, {Boquien}, {Burgarella}, {Salzer}, \& {Charlot}}]{2016Salim}
{Salim}, S., {Lee}, J.~C., {Janowiecki}, S., {et~al.} 2016, \apjs, 227, 2

\bibitem[{{Salpeter}(1955)}]{1955Salpeter}
{Salpeter}, E.~E. 1955, \apj, 121, 161

\bibitem[{{Schlegel} {et~al.}(1998){Schlegel}, {Finkbeiner}, \&
  {Davis}}]{1998Schlegel}
{Schlegel}, D.~J., {Finkbeiner}, D.~P., \& {Davis}, M. 1998, \apj, 500, 525

\bibitem[{{Shangguan} {et~al.}(2020){Shangguan}, {Ho}, {Bauer}, {Wang}, \&
  {Treister}}]{2020Shangguan}
{Shangguan}, J., {Ho}, L.~C., {Bauer}, F.~E., {Wang}, R., \& {Treister}, E.
  2020, \apjs, 247, 15

\bibitem[{{Shangguan} {et~al.}(2018){Shangguan}, {Ho}, \&
  {Xie}}]{2018Shangguan}
{Shangguan}, J., {Ho}, L.~C., \& {Xie}, Y. 2018, \apj, 854, 158

\bibitem[{{Shao} {et~al.}(2010){Shao}, {Lutz}, {Nordon}, {Maiolino},
  {Alexander}, {Altieri}, {Andreani}, {Aussel}, {Bauer}, {Berta},
  {Bongiovanni}, {Brandt}, {Brusa}, {Cava}, {Cepa}, {Cimatti}, {Daddi},
  {Dominguez-Sanchez}, {Elbaz}, {F{\"o}rster Schreiber}, {Geis}, {Genzel},
  {Grazian}, {Gruppioni}, {Magdis}, {Magnelli}, {Mainieri}, {P{\'e}rez
  Garc{\'\i}a}, {Poglitsch}, {Popesso}, {Pozzi}, {Riguccini}, {Rodighiero},
  {Rovilos}, {Saintonge}, {Salvato}, {Sanchez Portal}, {Santini}, {Sturm},
  {Tacconi}, {Valtchanov}, {Wetzstein}, \& {Wieprecht}}]{2010Shao}
{Shao}, L., {Lutz}, D., {Nordon}, R., {et~al.} 2010, \aap, 518, L26

\bibitem[{{Sil'chenko} \& {Moiseev}(2020)}]{2020Silchenko}
{Sil'chenko}, O., \& {Moiseev}, A. 2020, \aap, 638, L10

\bibitem[{Silk(2013)}]{Silk_2013}
Silk, J. 2013, The Astrophysical Journal, 772, 112

\bibitem[{{Skrutskie} {et~al.}(2006){Skrutskie}, {Cutri}, {Stiening},
  {Weinberg}, {Schneider}, {Carpenter}, {Beichman}, {Capps}, {Chester},
  {Elias}, {Huchra}, {Liebert}, {Lonsdale}, {Monet}, {Price}, {Seitzer},
  {Jarrett}, {Kirkpatrick}, {Gizis}, {Howard}, {Evans}, {Fowler}, {Fullmer},
  {Hurt}, {Light}, {Kopan}, {Marsh}, {McCallon}, {Tam}, {Van Dyk}, \&
  {Wheelock}}]{2006Skrutskie}
{Skrutskie}, M.~F., {Cutri}, R.~M., {Stiening}, R., {et~al.} 2006, \aj, 131,
  1163

\bibitem[{{Smethurst} {et~al.}(2021){Smethurst}, {Simmons}, {Coil}, {Lintott},
  {Keel}, {Masters}, {Glikman}, {Leung}, {Shanahan}, \&
  {Garland}}]{2021arXiv210805361S}
{Smethurst}, R.~J., {Simmons}, B.~D., {Coil}, A., {et~al.} 2021, arXiv
  e-prints, arXiv:2108.05361.
\newblock \doarXiv{2108.05361}

\bibitem[{{Smith} {et~al.}(2002){Smith}, {Tucker}, {Kent}, {Richmond},
  {Fukugita}, {Ichikawa}, {Ichikawa}, {Jorgensen}, {Uomoto}, {Gunn}, {Hamabe},
  {Watanabe}, {Tolea}, {Henden}, {Annis}, {Pier}, {McKay}, {Brinkmann}, {Chen},
  {Holtzman}, {Shimasaku}, \& {York}}]{2002Smith}
{Smith}, J.~A., {Tucker}, D.~L., {Kent}, S., {et~al.} 2002, \aj, 123, 2121

\bibitem[{{Stern} {et~al.}(2012){Stern}, {Assef}, {Benford}, {Blain}, {Cutri},
  {Dey}, {Eisenhardt}, {Griffith}, {Jarrett}, {Lake}, {Masci}, {Petty},
  {Stanford}, {Tsai}, {Wright}, {Yan}, {Harrison}, \& {Madsen}}]{2012Smid-ir}
{Stern}, D., {Assef}, R.~J., {Benford}, D.~J., {et~al.} 2012, \apj, 753, 30

\bibitem[{{Tempel} {et~al.}(2011){Tempel}, {Saar}, {Liivam{\"a}gi}, {Tamm},
  {Einasto}, {Einasto}, \& {M{\"u}ller}}]{2011Tempel}
{Tempel}, E., {Saar}, E., {Liivam{\"a}gi}, L.~J., {et~al.} 2011, \aap, 529, A53

\bibitem[{{Tolman}(1930)}]{1930PTolman}
{Tolman}, R.~C. 1930, Proceedings of the National Academy of Science, 16, 511

\bibitem[{{Trujillo} {et~al.}(2001){Trujillo}, {Aguerri}, {Cepa}, \&
  {Guti{\'e}rrez}}]{2001Trujilo}
{Trujillo}, I., {Aguerri}, J.~A.~L., {Cepa}, J., \& {Guti{\'e}rrez}, C.~M.
  2001, \mnras, 328, 977

\bibitem[{van~de Voort {et~al.}(2016)van~de Voort, Bahé, Bower, Correa, Crain,
  Schaye, \& Theuns}]{gasaccretion2016}
van~de Voort, F., Bahé, Y.~M., Bower, R.~G., {et~al.} 2016, \mnras, 466, 3460

\bibitem[{{Wang} {et~al.}(2012){Wang}, {Kauffmann}, {Overzier}, {Tacconi},
  {Kong}, {Saintonge}, {Catinella}, {Schiminovich}, {Moran}, \&
  {Johnson}}]{Wangetal2012}
{Wang}, J., {Kauffmann}, G., {Overzier}, R., {et~al.} 2012, \mnras, 423, 3486

\bibitem[{{Weistrop} {et~al.}(1995){Weistrop}, {Hintzen}, {Liu}, {Lowenthal},
  {Cheng}, {Oliversen}, {Brown}, \& {Woodgate}}]{1993Weistrop}
{Weistrop}, D., {Hintzen}, P., {Liu}, C., {et~al.} 1995, \aj, 109, 981

\bibitem[{Williams \& Evans(2017)}]{flatbar}
Williams, A.~A., \& Evans, N.~W. 2017, \mnras, 469, 4414

\bibitem[{{Wright} {et~al.}(2010){Wright}, {Eisenhardt}, {Mainzer}, {Ressler},
  {Cutri}, {Jarrett}, {Kirkpatrick}, {Padgett}, {McMillan}, {Skrutskie},
  {Stanford}, {Cohen}, {Walker}, {Mather}, {Leisawitz}, {Gautier}, {McLean},
  {Benford}, {Lonsdale}, {Blain}, {Mendez}, {Irace}, {Duval}, {Liu}, {Royer},
  {Heinrichsen}, {Howard}, {Shannon}, {Kendall}, {Walsh}, {Larsen}, {Cardon},
  {Schick}, {Schwalm}, {Abid}, {Fabinsky}, {Naes}, \& {Tsai}}]{2010AWise}
{Wright}, E.~L., {Eisenhardt}, P. R.~M., {Mainzer}, A.~K., {et~al.} 2010, \aj,
  140, 1868

\bibitem[{Yi {et~al.}(2011)Yi, Lee, Sheen, Jeong, Suh, \& Oh}]{Yi_2011}
Yi, S.~K., Lee, J., Sheen, Y.-K., {et~al.} 2011, \apjs, 195, 22

\bibitem[{{York} {et~al.}(2000){York}, {Adelman}, {Anderson}, {Anderson},
  {Annis}, {Bahcall}, {Bakken}, {Barkhouser}, {Bastian}, {Berman}, {Boroski},
  {Bracker}, {Briegel}, {Briggs}, {Brinkmann}, {Brunner}, {Burles}, {Carey},
  {Carr}, {Castander}, {Chen}, {Colestock}, {Connolly}, {Crocker}, {Csabai},
  {Czarapata}, {Davis}, {Doi}, {Dombeck}, {Eisenstein}, {Ellman}, {Elms},
  {Evans}, {Fan}, {Federwitz}, {Fiscelli}, {Friedman}, {Frieman}, {Fukugita},
  {Gillespie}, {Gunn}, {Gurbani}, {de Haas}, {Haldeman}, {Harris}, {Hayes},
  {Heckman}, {Hennessy}, {Hindsley}, {Holm}, {Holmgren}, {Huang}, {Hull},
  {Husby}, {Ichikawa}, {Ichikawa}, {Ivezi{\'c}}, {Kent}, {Kim}, {Kinney},
  {Klaene}, {Kleinman}, {Kleinman}, {Knapp}, {Korienek}, {Kron}, {Kunszt},
  {Lamb}, {Lee}, {Leger}, {Limmongkol}, {Lindenmeyer}, {Long}, {Loomis},
  {Loveday}, {Lucinio}, {Lupton}, {MacKinnon}, {Mannery}, {Mantsch}, {Margon},
  {McGehee}, {McKay}, {Meiksin}, {Merelli}, {Monet}, {Munn}, {Narayanan},
  {Nash}, {Neilsen}, {Neswold}, {Newberg}, {Nichol}, {Nicinski}, {Nonino},
  {Okada}, {Okamura}, {Ostriker}, {Owen}, {Pauls}, {Peoples}, {Peterson},
  {Petravick}, {Pier}, {Pope}, {Pordes}, {Prosapio}, {Rechenmacher}, {Quinn},
  {Richards}, {Richmond}, {Rivetta}, {Rockosi}, {Ruthmansdorfer}, {Sandford},
  {Schlegel}, {Schneider}, {Sekiguchi}, {Sergey}, {Shimasaku}, {Siegmund},
  {Smee}, {Smith}, {Snedden}, {Stone}, {Stoughton}, {Strauss}, {Stubbs},
  {SubbaRao}, {Szalay}, {Szapudi}, {Szokoly}, {Thakar}, {Tremonti}, {Tucker},
  {Uomoto}, {Vanden Berk}, {Vogeley}, {Waddell}, {Wang}, {Watanabe},
  {Weinberg}, {Yanny}, {Yasuda}, \& {SDSS Collaboration}}]{2000York}
{York}, D.~G., {Adelman}, J., {Anderson}, John~E., J., {et~al.} 2000, \aj, 120,
  1579

\bibitem[{{Zhang} {et~al.}(2021){Zhang}, {Peng}, {Ho}, {Maiolino}, {Renzini},
  {Mannucci}, {Dekel}, {Guo}, {Li}, {Yuan}, {Lilly}, {Dou}, {Guo}, {Man}, {Li},
  \& {Shi}}]{2021Zhang}
{Zhang}, C., {Peng}, Y., {Ho}, L.~C., {et~al.} 2021, \apj, 911, 57

\bibitem[{{Zhong} {et~al.}(2008){Zhong}, {Liang}, {Liu}, {Hammer}, {Hu},
  {Chen}, {Deng}, \& {Zhang}}]{2008Zhong}
{Zhong}, G.~H., {Liang}, Y.~C., {Liu}, F.~S., {et~al.} 2008, \mnras, 391, 986

\end{thebibliography}
\bibliographystyle{aasjournal}



%{\color{red}KS: the following para can be in the discussion.}-Done

%\citet{2009kannappan} suggests that most of blue early type high mass galaxies (log($M_\star$/\msun) $\sim 10$) are likely a major-merger remnants that are fading to red-sequence. Although, it highly unlikely to witness a major-merger event in voids \citep{SJPenny,2021Florez}, we plan to do structural decomposition of the galaxy to look for any major-merger signatures in the morphology and if the merger interaction is ruled-out, we investigate the source of the observed color in the galaxy.  

%{\color{red}KS: This has already been said}-Done

%{\color{red}KS: This part could go above as motivational argument.}-DOne 

%{\color{red}KS: The following is not introduction but some sort of work summary.}-Done

%As we are dealing with an early-type galaxy with no hint of spiral arms in its morphology, we would deduce if the emission from the galaxy is signature of UV upturn \citep{OConnell1999} or recent star-formation \citep{kennicutt2012}. In addition, 

\end{document}